\documentclass[11pt]{article}
\usepackage[latin1]{inputenc}
\usepackage{amsfonts}
\usepackage{amssymb,color}
\usepackage{graphicx}
\newcommand{\be}{\begin{equation}}
\newcommand{\ee}{\end{equation}}
\newcommand{\ba}{\begin{array}}
\newcommand{\ea}{\end{array}}
\newcommand{\bea}{\begin{eqnarray}}
\newcommand{\eea}{\end{eqnarray}}
\newcommand{\bee}{\begin{eqnarray*}}
\newcommand{\eee}{\end{eqnarray*}}

\catcode`@=11
\renewcommand\appendix{\bigskip {\noindent\Large \bf Appendix}\par
  \setcounter{section}{0}%
  \setcounter{subsection}{0}%
  \renewcommand\thesection{\@Alph\c@section}}
\catcode`@=12

\def\Section{\setcounter{equation}{0}\section}
\newtheorem{theorem}{Theorem}[section]

\newtheorem{proposition}[theorem]{Proposition}

\def\thesection{\arabic{section}}

\newenvironment{acknowledgement}{\noindent{\bf Acknowledgement.~}}{}

\def\RR{\mathbb{R}}
\def\R{{\mathbb R}}
\def\N{{\mathbb N}}
\def\S{{\mathbb S}}
\def\fc{{\hat f}}
\def\lim{\mathop{\rm lim}}

\def\exp{{\rm exp}}

\def\fref#1{{\rm (\ref{#1})}}

\def\ds{\displaystyle}

\def\ds{\displaystyle}

\title{Escape of stars  from  gravitational clusters in the Chandrasekhar model}
\date{}

\author{Mohammed Lemou$^{*,****}$ and Pierre-Henri Chavanis $^{**,***}$}

\date{\it $^{*}$ Universit\'e Rennes 1,  IRMAR, F-35042, France\\
$^{**}$ Universit\'e de Toulouse, UPS, Laboratoire de  Physique Th\'eorique (IRSAMC), F-31062, France \\
$^{***}$ CNRS,  Laboratoire de  Physique Th\'eorique (IRSAMC), F-31062, France\\
$^{****}$ CNRS,  Institut de recherche math\'ematique de Rennes (IRMAR), F-35042, France}

\begin{document}

\maketitle

%%%%%%%%%%%%%%%%%%%%%%%%%%%%%%%%%%%%%%%%%%%%%%%%%%%%%
%%%%%%%%%%%%%%%%%%%%%%%%%%%%%%%%%%%%%%%%%%%%%%%%%%%%%
\begin{abstract}

We study the evaporation of stars from globular clusters using the
simplified Chandrasekhar model [S. Chandrasekhar, Astrophys. J.  {\bf
97}, 263 (1943)]. This is an analytically tractable model giving
reasonable agreement with more sophisticated models that require
complicated numerical integrations. In the Chandrasekhar model: (i)
the stellar system is assumed to be infinite and homogeneous (ii) the
evolution of the velocity distribution of stars $f(v,t)$ is governed
by a Fokker-Planck equation, the so-called Kramers-Chandrasekhar
equation (iii) the velocities $|v|$ that are above a threshold value
$R>0$ (escape velocity) are not counted in the statistical
distribution of the system. In fact, high velocity stars leave the
system, due to free evaporation or to the attraction of a neighboring
galaxy (tidal effects). Accordingly, the total mass and energy of the
system decrease in time. If the star dynamics is described by the
Kramers-Chandrasekhar equation, the mass decreases to zero
exponentially rapidly. Our goal is to obtain {\it non-perturbative}
analytical results that complement the seminal studies of
Chandrasekhar, Michie and King valid for large times $t\rightarrow
+\infty$ and large escape velocities $R\rightarrow +\infty$.  In
particular, we obtain an exact semi-explicit solution of the
Kramers-Chandrasekhar equation with the absorbing boundary condition
$f(R,t)=0$. We use it to obtain an explicit expression of the mass
loss at any time $t$ when $R\rightarrow +\infty$. We also derive an
exact integral equation giving the exponential evaporation rate
$\lambda(R)$, and the corresponding eigenfunction $f_{\lambda}(v)$,
when $t\rightarrow +\infty$ for any sufficiently large value of the
escape velocity $R$. For $R\rightarrow +\infty$, we obtain an explicit
expression of the evaporation rate that refines the Chandrasekhar
results. More generally, our results can have applications in other
contexts where the Kramers equation applies, like the classical
diffusion of particles over a barrier of potential (Kramers problem).

\end{abstract}

%%%%%%%%%%%%%%%%%%%%%%%%%%%%%%%%%%%%%%%%%%%%%%%%%%%%%
%%%%%%%%%%%%%%%%%%%%%%%%%%%%%%%%%%%%%%%%%%%%%%%%%%%%%

%

%%%%%%%%%%%%%%%%%%%%%%%%%%%%%%%%%%%%%%%%%%%%%
%%%%%%%%%%%%%%%%%%%%%%%%%%%%%%%%%%%%%%%%%%%%%

\Section{Introduction}

In a seminal paper, Chandrasekhar \cite{chandra} developed a Brownian
theory of stellar dynamics in order to determine the rate of escape of
stars from globular clusters. Small groups of stars tend to approach a
statistical equilibrium state (described by the Boltzmann
distribution) as a result of stellar encounters. However, high energy
stars are not bound to the system and escape to infinity. For an
isolated system, the average escape velocity for all stars in the
cluster is fixed by the virial theorem according to the equation
$R=2v_{rms}$ where $v_{rms}=\langle v^2\rangle^{1/2}$ is the
root-mean-square velocity (RMS) \cite{bt}.  If the system, e.g. a
globular cluster, is submitted to tidal forces from a neighboring
galaxy, the escape velocity can be smaller.  Therefore, stars clusters
tend to slowly evaporate. This evaporation was first studied by
Ambartsumian \cite{ambartsumian} and Spitzer \cite{spitzer} using
phenomenological arguments. They estimated the evaporation time by
removing a fraction $\gamma=7.38\ 10^{-3}$ of stars every relaxation
time, where $\gamma$ is the fraction of particles in a Maxwellian
distribution that have speeds exceeding twice the RMS velocity.  In a
more precise treatment, Chandrasekhar \cite{chandra} described the
``collisional'' evolution of a stellar system by a Fokker-Planck
equation (the nowadays called Kramers equation) involving a diffusion
term in velocity space modeling the erratic motion of the stars and a
friction term that appears to be necessary to drive the system towards
the Boltzmann distribution predicted by statistical mechanics
(fluctuation-dissipation theorem). The diffusion coefficient and the
``dynamical friction", satisfying the Einstein relation, were
independently justified from kinetic theory by explicitly calculating
the first and second moments of the velocity increment suffered by a
star during a succession of binary encounters. In order to account for
evaporation, Chandrasekhar imposed as a boundary condition that the
distribution function $f(v,t)$ vanishes when the star velocity reaches
a maximum value $|v|=R$. He then reduced the problem to the study of
an eigenvalue equation in a bounded domain of velocities $|v|\le
R$. The fundamental eigenvalue gives the exponential evaporation rate
of the stars from the cluster for $t\rightarrow +\infty$ and the
associated eigenfunction gives the quasi-stationary distribution
function of the system. This distribution function is close to the
Boltzmann distribution for low velocities and tends to zero at the
escape velocity. Chandrasekhar solved the eigenvalue problem by
transforming the Kramers equation into a Schr\"odinger equation (with
imaginary time) for a quantum oscillator in a box and by expanding the
solutions of that equation in the form of a series. He obtained
(semi-explicit) analytical results in the $R\rightarrow +\infty$ limit
or, equivalently, for a small evaporation rate $\lambda(R)\rightarrow
0$. In his first treatment
\cite{chandra}, he assumed that the diffusion coefficient is constant
and in a more exact theory \cite{chandramore}, he took into account
the dependence of the diffusion coefficient with the velocity.  His
work was followed by Spitzer \& H\"arm \cite{sh} who determined the
escape rate (eigenvalue) and the quasi-stationary distribution
function (eigenfunction) numerically for any value of the escape
velocity $R$.  Then, Michie \cite{michie} and King \cite{king}
obtained for $R\rightarrow +\infty$ a simple analytical expression of
the quasi-stationary distribution function in the form of a lowered
isothermal distribution which vanishes at the escape velocity. This
leads to the so-called Michie-King model \cite{bt} that is
asymptotically valid in the limit $R\rightarrow +\infty$.

The Chandrasekhar model described previously is based on simplifying
assumptions. It is first assumed that the system is spatially
homogeneous and infinite while globular clusters are highly
inhomogeneous and limited in space. On the other hand, the collisional
evolution of the system is modeled by the Kramers equation while a
more relevant equation is the gravitational Landau equation that is
the standard kinetic equation of stellar dynamics
\cite{bt,spitzerbook,paddy,epjb}.  The Kramers equation corresponds to a
canonical description in which the system is assumed to be in contact
with a thermal bath from which it can extract energy so that the
temperature $T$ is fixed.  Alternatively, the Landau equation
corresponds to a microcanonical description in which the system is
assumed to be isolated so that the energy $E$ is conserved.  Since
globular clusters are isolated Hamiltonian systems (up to the slow
evaporation process), the microcanonical description appears to be
more relevant. Therefore, when we take into account spatial
inhomogeneity and model the encounters in a self-consistent way, the
proper model to consider is formed by the gravitational Landau
equation coupled to the Poisson equation.  In order to go beyond the
limitations of the Chandrasekhar model and obtain more accurate rates
of escape, the astrophysicists have performed numerical simulations of
stellar systems. Different types of simulations have been
performed. They solved (i) the $N$-body Hamiltonian problem associated
to the Newton equations \cite{al} (ii) the hydrodynamic moments of the
Landau equation \cite{larson,lbe} (iii) the $N$-body problem where the
effect of encounters is modeled by Monte Carlo methods
\cite{henonmc,spitzermc,shapiro}, and (iv) the orbit-averaged-Fokker-Planck equation
\cite{cohn}. These methods are reviewed in the books of Spitzer
\cite{spitzerbook} and Binney \& Tremaine \cite{bt} and in the 
reviews \cite{mh,ls}. In these numerical works, the spatial
inhomogeneity of the cluster is properly taken into account.

These simulations have led to the following scenario for the evolution
of globular clusters \cite{bt,spitzerbook}. In a first regime, a
self-gravitating system initially out-of-mechanical equilibrium
undergoes a process of {\it violent collisionless relaxation} towards
a virialized state\footnote{This form of relaxation is appropriate to
account for the actual structure of elliptical galaxies whose dynamics is
encounterless for the timescales of interest \cite{bt}.}. In this regime, the
dynamical evolution of the cluster is described by the Vlasov-Poisson
system and the phenomenology of violent relaxation has been described
by H\'enon \cite{henonvr}, King \cite{kingvr} and Lynden-Bell
\cite{lb}. Numerical simulations that start from cold and clumpy
initial conditions generate a quasi stationary state (QSS) that fits
the de Vaucouleurs $R^{1/4}$ law quite well \cite{vanalbada}. The
inner core is almost isothermal (as predicted by Lynden-Bell
\cite{lb}) while the velocity distribution in the envelope 
is radially anisotropic and the density profile decreases like
$r^{-4}$ \cite{sb,hm}. One success of Lynden-Bell's statistical theory
of violent relaxation is to explain the isothermal core without
recourse to ``collisions''. By contrast, the structure of the halo
cannot be explained by Lynden-Bell's theory as it is a result of an
{\it incomplete relaxation}. On longer timescales, encounters between
stars must be taken into account and the dynamical evolution of the
cluster is governed by the Vlasov-Landau-Poisson system which is the
standard model of stellar dynamics. This collisional regime is
appropriate to understand the actual structure of globular
clusters. In this regime, the system passes through a succession of
quasi stationary states (QSS) that are steady states of the Vlasov
equation slowly evolving in time due to the cumulative effect of
encounters. The first stage of the collisional evolution is driven by
{\it evaporation}. Due to a series of weak encounters, the energy of a
star can gradually increase until it reaches the local escape energy;
in that case, the star leaves the system\footnote{There can also be a
process of {\it ejection}
\cite{henonejection} in which a single close encounter produces a
velocity change that is sufficient to eject the star out of the
cluster. However, it can be shown that this process is less efficient
than evaporation.}. Numerical simulations \cite{spitzerbook} show that
during this regime the system reaches a quasi-stationary state that
slowly evolves in amplitude due to evaporation as the system loses
mass and energy. This quasi stationary distribution function (DF) is
close to the Michie-King model. The system has a core-halo
structure. The core is isothermal while the stars in the outer halo
move in predominantly radial orbits. Therefore, the distribution
function in the halo is anisotropic.  The density follows the
isothermal law $\rho\sim r^{-2}$ in the central region (with a core of
almost uniform density) and decreases like $\rho\sim r^{-7/2}$ in the
halo \cite{bt}. Due to evaporation, the halo expands while the core
shrinks as required by energy conservation. At some point of the
evolution, when the energy passes below a critical value (or when the
density contrast becomes sufficiently high), the system undergoes an
instability related to the Antonov \cite{antonov} instability and the
gravothermal catastrophe takes place
\cite{lbw}. This instability is due to the negative specific heat of
the inner system that evolves by losing energy and thereby growing
hotter (see reviews in \cite{paddy,revue}). This leads to {\it core
collapse} \cite{bt}. Mathematically speaking, core collapse would
generate a finite time singularity: if the evolution is modeled by the
orbit-averaged-Fokker-Planck equation, Cohn \cite{cohn} finds that the
collapse is self-similar, that the central density becomes infinite in
a finite time and that the density behaves like $\rho\sim r^{-2.23}$
(if the evolution is modeled by the Landau-Poisson system, it is
argued in \cite{lk} that the density behaves like $\rho\propto r^{-3}$
in the final stage of the collapse). In reality, if we come back to
the $N$-body system, core collapse is arrested by the formation of
binary stars. These binaries can release sufficient energy to stop the
collapse
\cite{henonbinary} and even drive a re-expansion of the cluster in a
post-collapse regime
\cite{inagakilb}. Then, in principle, a series of gravothermal
oscillations should follow
\cite{bettwieser}. In practice, the processes of evaporation and core
collapse take place simultaneously so that it is difficult to isolate
the effect of any single process in the evolution of a globular
cluster.

Concerning the evaporation process, the Princeton
code was the first code to yield reliable evaporation rates
\cite{princeton} giving $t_{evap}=-N(dN/dt)^{-1}\simeq 300 t_{rh}$ for
isolated clusters\footnote{Tidal forces from the Galaxy can increase
the evaporation rate \cite{chevalier}.}. These results are not very
different from those obtained with the spatially homogeneous
Kramers-Chandrasekhar equation.  Our goal in this paper is not to make
a realistic modeling of stellar systems but rather to consider simple
models of evaporation that are {\it analytically
tractable}. Therefore, we shall use the Chandrasekhar model which
yields a reasonable description of the evaporation process in globular
clusters and which can be studied analytically. Chandrasekhar solved
the problem perturbatively: he first considered the long time limit
$t\rightarrow +\infty$ so that the distribution function $f_R(v,t)$ is
dominated by the contribution of the fundamental eigenmode
$f_R(v)e^{-|\lambda(R)| t}$ and then took the limit $R\rightarrow
+\infty$ to obtain an approximate expression of the quasi-stationary
distribution $f_R(v)$ (fundamental eigenfunction) and escape rate
$\lambda(R)$ (fundamental eigenvalue). In this paper, we shall
reconsider the Chandrasekhar problem on a new angle which allows to
obtain {\it non-perturbative} results. In particular, we find an exact
semi-explicit solution of the Kramers equation with boundary condition
$f(v,t)=0$ when $|v|=R$. This solution $f(v,t)$ depends on the
remaining mass in the cluster $M_0(t)$ which satisfies an autonomous
equation.  We use this general formula to obtain: (i) the mass
$M_0(t)$ for any fixed time $t$ in the limit $R\rightarrow +\infty$,
(ii) an exact integral equation for the eigenvalue $\lambda(R)$ of the
fundamental mode (evaporation rate) valid for any sufficiently large
$R$, (iii) an exact explicit expression of the fundamental
eigenfunction valid for any sufficiently large $R$, and (iv) an
explicit asymptotic expression of the evaporation rate when
$R\rightarrow +\infty$. Therefore, our approach complements
Chandrasekhar's original work and offers new perspectives. Our main
results are, however, restricted to the Kramers equation, i.e. a
Fokker-Planck equation with constant diffusion coefficient and
quadratic potential (linear friction). A different approach that goes
beyond these limitations (but which is restricted to the asymptotic
limits $t\rightarrow +\infty$ and $R\rightarrow +
\infty$) is developed in Appendix \ref{sec_comp}.

Let us finally note that our approach is not limited to the
astrophysical problem mentioned above but that it can have applications in
different area.  First of all, Chandrasekhar's study of the rate of
escape of stars from globular clusters is closely connected to the
classical Kramers \cite{kramers} problem for the escape rate of a
Brownian particle across a potential barrier that has many
applications in physics and chemistry (surprisingly, Chandrasekhar
\cite{chandra} did not mention this connection).  In that case, the
problem is usually formulated in $d=1$ dimension. On the other hand,
Chandrasekhar's procedure has been used in the context of planet
formation \cite{vortex} in order to determine the rate of escape of
dust from large-scale vortices (assumed to be present in the solar
nebula) due to turbulence. In that case, the problem is
two-dimensional. In view of the fundamental nature of the mathematical problem, it
seems relevant to develop our formalism in arbitrary dimension of
space $d$ in order to cover a wide range of possible applications.

%%%%%%%%%%%%%%%%%%%%%%%%%%%%%%%%%%%%%%%%%%%%%
%%%%%%%%%%%%%%%%%%%%%%%%%%%%%%%%%%%%%%%%%%%%%

\section{Setting of the problem and statement of the results}
\label{sec_problem}

%%%%%%%%%%%%%%%%%%%%%%%%%%%%%%%%%%%%%%%%%%%%%
%%%%%%%%%%%%%%%%%%%%%%%%%%%%%%%%%%%%%%%%%%%%%

\subsection{Kinetic models on  a bounded  velocity domain}
\label{sec_models}

Basically, the evolution of the distribution function $f(x,v,t)$ of a
stellar system is described by the Vlasov-Landau-Poisson equation \cite{bt,spitzerbook,paddy,epjb}:
\be\label{Landau}
\left   \{ \begin{array}{ll}
         \ds \frac{\partial f}{\partial t}+v\cdot \nabla_x f-\nabla\Phi\cdot \nabla_v f =\nabla_v\cdot\int K (v-v_*) \left(
         f(x,v_*)\nabla_v f(x,v) - f(x,v)\nabla_v f(x,v_*)\right) dv_*, \\ \ds
        \Delta\Phi=4\pi G\int f\, dv, \\  \ds
   f(x,v,t=0)=f_0(x,v)\geq 0, \ \ (x,v)\in \RR^3\times \RR^3,\\ \ds
         \end{array}
\right.
\end{equation}
where $\Phi(x,t)$ is the gravitational potential and $K(u)$ is the
following $3\times 3$  matrix
\be\label{Phi}
 K(u)_{ij} = A \frac{ |u|^2 \delta_{ij} - u_iu_j}{ |u|^{3}},
\end{equation}
where $A=2\pi G^2 m\ln\Lambda$ is a constant ($G$ is the gravity
constant, $m$ the mass of a star and $\ln\Lambda$ the Coulomb
logarithm). These equations must be complemented by the boundary
condition $f(x,v,t)=0$ if $\epsilon\equiv v^2/2+\Phi(x,t)>0$ which
expresses the fact that stars with positive energy are lost by the
system. Indeed, they are unbound and free to escape to infinity. This
is the reason for the evaporation of the star cluster. The
Vlasov-Landau-Poisson system (\ref{Landau}) is the standard model of
stellar dynamics. From the Landau equation, the relaxation time due to two-body
encounters can be estimated by $t_R\sim (N/\ln N)t_D$ where $t_D$ is
the dynamical time and $N$ the number of stars in the system
\cite{bt}. For large groups of stars like elliptical galaxies ($N\sim
10^{11}$, $t_D\sim 10^8$ years, age $\sim 10^8$ years), the relaxation
time is much larger than the age of the universe by several orders of
magnitude so that star encounters are completely negligible. In that
case, their evolution is described by the Vlasov-Poisson system,
i.e. by (\ref{Landau}) with the r.h.s. taken equal to zero.  For
smaller groups of stars like globular clusters ($N\sim 10^{5}$,
$t_D\sim 10^5$ years, age $\sim 10^{10}$ years) whose ages are of the
same order as the relaxation time, the encounters must be taken into
account.  The study of the full Vlasov-Landau-Poisson equation is
extremely complicated because it involves several processes: (i)
violent relaxation in the collisionless regime (ii) gravothermal
catastrophe in the collisional regime, and  (iii) evaporation. Furthermore,
the coupling between the Landau equation and the Poisson equation, and
the fact that the distribution function depends on seven variables
$(x,v,t)$ in the general case, make these equations untractable
without further assumptions. Some simplification can be obtained by
averaging over the orbits of the stars thereby obtaining the
orbit-averaged-Fokker-Planck equation
\cite{bt,spitzerbook}. For spherical systems, this leads to an equation
for the distribution function $f=f(\epsilon,t)$ that depends only on
the energy $\epsilon=v^2/2+\Phi$ and time $t$. Still, the theoretical
study of the orbit-averaged-Fokker-Planck equation remains very
complicated. In order to distinguish the contribution of each process
occurring in the evolution of a stellar system and to study
specifically the evaporation process in a very simple setting (which
is our motivation here) we shall make additional simplifying
assumptions. First, we shall disregard the spatial structure of the
system and assume that the medium under consideration is infinite and
homogeneous. If we implement this approximation naively, solving the
spatially homogeneous Landau equation for $f(v,t)$, it is found
\cite{kingL} that the r.m.s. velocity decreases due to evaporation
while in reality (i.e. when spatial inhomogeneity is retained and the Landau equation is coupled to the Poisson equation) the
contraction of the core causes the r.m.s. velocity to increase (as
potential energy is converted into kinetic energy faster than the
escaping stars carry energy away). This approximation leads therefore
to unphysical results. This problem was solved by King
\cite{kingL} by adding artificially in the spatially homogeneous
Landau equation an additional outward flux in velocity space. In that
case, the decrease in kinetic energy due to evaporation is compensated
by the increase in kinetic energy due to contraction. Another
solution, that was proposed earlier by Chandrasekhar \cite{chandra},
is to assume that the star under consideration has encounters with a
separate group of stars having a fixed (usually assumed Maxwellian)
velocity distribution. In that case, the
star is able to extract energy from a reservoir imposing its
temperature (thermal bath). Therefore, Chandrasekhar models the
evolution of the system by a Fokker-Planck equation of the form\footnote{Chandrasekhar \cite{chandra} did not
give a precise justification for using a differential Fokker-Planck
equation instead of an integro-differential equation like the Landau
equation. The Fokker-Planck equation considered by Chandrasekhar
describes the evolution of a ``test star'' in a bath of ``field
stars'' at statistical equilibrium with fixed temperature (canonical
description). By contrast, the Landau equation describes the evolution
of the system as a whole and conserves the energy (microcanonical
description). Apparently, the Landau equation was not well-known by
the astrophysical community at that time (see discussion in
\cite{epjb}). Here, we shall assume that the existing contraction of
the cluster has the effect to ``heat up'' the stars and balance their
``cooling'' due to evaporation. As a result of these two antagonistic
effects, we can assume that the temperature (velocity dispersion) of
the stars remains fixed. Therefore, everything happens as if the
system were in contact with a thermal bath, justifying the
Chandrasekhar assumptions. Accordingly, when we disregard the spatial
structure of the system (but take it into account indirectly as
explained above), it is more appropriate to model the dynamics of the
star cluster by the Fokker-Planck equation (canonical) rather than by
the Landau equation (microcanonical). However, if we take into account the
spatial structure of the system, the best kinetic description is the
Landau equation coupled to the Poisson equation. }:
\be
\label{FP}
\left   \{ \begin{array}{ll}
         \ds \frac{\partial f}{\partial t}  (v,t) =
         Q_R^{FP}(f)(v)=\nabla\cdot \left[
         D(|v|)\left(  \nabla f(v) + \beta  f(v) v\right)\right],  \ v \in B_R, \\
         \ds f(v,t=0)=f_0(v)\geq 0, \ \ v\in B_R\\
          \ds f(v,t)=0, \ \ \ \mbox{if} \  |v|=R,
         \end{array}
\right.
\ee
where  $D(|v|)$  is  some  given  nonnegative   diffusion matrix.  The
Fokker-Planck    equation (\ref{FP}) can    be derived from the Landau
equation   (\ref{Landau})   by making   the   so-called ``thermal bath
approximation'',  i.e.   by  replacing    the  function  $f(v_*)$   in
(\ref{Landau}) by    the  Maxwell  distribution   $f(v)=  {\rho}{(2\pi
T)^{-3/2}} \exp  (-{v^2}/{2T})$  with  inverse temperature $\beta=1/T$
(here, the  mass of stars has been  included in the temperature). This
procedure transforms  an integrodifferential (Landau) kinetic equation
into a differential   (Fokker-Planck) equation and allows an  explicit
computation of the diffusion coefficient $D(|v|)$ (see \cite{epjb} and
references therein). The Fokker-Planck equation (\ref{FP}) can also be
derived from a  stochastic Langevin  equation  with a  linear friction
(Ornstein-Uhlenbeck  process)  and is   called  the Kramers   equation
\cite{kramers}. Since  it was  derived independently  by Chandrasekhar
\cite{chandra} in the astrophysical context, we  will call it here the
Kramers-Chandrasekhar equation.  Finally, the last condition in
(\ref{FP}) expresses the fact that a star with velocity above a
certain threshold $R$ escapes and is therefore lost by the system.
This threshold can be estimated by the following classical argument
\cite{bt}. The escape speed at $x$ is given by $v_e^2=-2\Phi(x)$,
corresponding to $\epsilon=0$. The mean square escape speed in a
system whose density is $\rho(x)$ is therefore $\langle
v_e^2\rangle=\int \rho(x)v_e^2\, dx/\int \rho(x)\, dx= -(2/M)\int
\rho(x)\Phi(x)\, dx=-4W/M$ where $M$ is the total mass and $W$ the
potential energy. According to the virial theorem $-W=2K$ where
$K=(1/2)M\langle v^2\rangle$ is the kinetic energy.  Hence $\langle
v_e^2\rangle=4\langle v^2\rangle$.  For a Maxwellian distribution with
temperature $T$, we obtain $R=2\sqrt{3T}$. However this is just an
estimate. Furthermore, the escape velocity can be smaller if the
system (globular cluster) is subject to the tide of a nearby galaxy.
Therefore, for sake of generality, we shall consider $R$ arbitrary.
The Fokker-Planck equation (\ref{FP}) with proper boundary condition
$f(R,t)=0$ was used by Chandrasekhar \cite{chandra} and others
\cite{sh,michie,king} to determine the rate of escape of stars from
globular clusters.  We shall call it the {\it Chandrasekhar
model}. This is the model studied in the present paper.  For sake of
generality, we shall consider these equations in $d$ dimensions.

%%%%%%%%%%%%%%%%%%%%%%%%%%%%%%%%%%
%%%%%%%%%%%%%%%%%%%%%%%%%%%%%%%%%

\subsection{Semi-explicit solutions  on a bounded velocity domain}

In this section, we focus on the Fokker-Planck equation \fref{FP} with $D\equiv 1$ and
boundary condition $f(R,t)=0$. We prove
that a semi-explicit solution of this kinetic equation can be given for
this model.  More precisely, we derive an autonomous relation
satisfied by $M_0(t)$, the mass remaining in the cluster at time $t$, and then give an explicit expression of the solution $f(v,t)$ in terms of $M_0(t)$ and of the initial data
$f_0(v)$ only. This result will be used to determine the exact (i.e. non perturbative) rate of
escape of stars in the Chandrasekhar model (see section
2.3).  

Note that the difficulty here comes from the
boundedness character of the velocity domain and the
presence of the term $v f$ in the drift-diffusion model
\fref{FP}, which prevent a direct use of
Fourier analysis.  Our strategy is to first transform Eq. \fref{FP}
into equivalent equations {\em on the whole domain $\R^d$} where the
boundary condition naturally disappears and is replaced by a source
term (see Appendix \ref{sec_a1}). The resulting equation makes sense
in the space of distributions and one can use Fourier techniques in
this space to work out a semi-explicit solution of the problem.

\begin{proposition}[Semi-explicit solution of  \fref{FP}]
\label{prop2}
Let $f_0$ be a smooth and isotropic initial data supported inside
$B_R$. The solution to \fref{FP} when $D\equiv 1$, with initial data $f_0$,   is given by
\be
\label{solution-exacte-d}
\begin{array}{ll} f(v,t) = & \ds \frac{1}{(2\pi)^d}
  \left(\frac{\pi}{A(0,t)}\right)^{d/2} \int_{B_R} f_0(v_*)
  \ \exp\left(-\frac{|v- B(0,t)v_*|^2}{4A(0,t)}\right) dv_*\  \\
 & \ds {+}\frac{1}{(2\pi)^d} \frac{1}{|\S_1^d|}\int_0^t ds
  \left(\frac{\pi}{A(s,t)}\right)^{d/2} M_0'(s)\int_{\sigma \in \S_1^d}
  d\sigma \ \exp\left(-\frac{|B(s,t)R\sigma +v|^2}{4A(s,t)}\right),
\end{array}
\ee
where $\S_1^d$ is the unit sphere $\R^d$ with measure $|\S_1^d|$,  $d\sigma$  is the surface element of this unit sphere, and:
$$A(s,t)= \frac{1}{2\beta} \left( 1- \exp( -2\beta (t-s))\right), \ \ \
\  B(s,t)= \exp(-\beta(t-s)).$$
The total mass $M_0(t)$  of $f$,
defined by  
\be\label{M0M1}
M_0(t)= \int_{B_R}  f(v,t)\, dv, 
\ee
is determined  from the boundary condition   $f(R\omega,t)=0, \ \ \forall \omega \in \S_1^d$ leading to the autonomous equation
\be
\label{equation-masse}
\begin{array}{ll}
  \ds A(0,t)^{-d/2} &\ds  \int_{B_R} f_0(v_*)
  \exp\left(-\frac{|R\omega-B(0,t)v_*|^2}{4A(0,t)}\right) dv_*  \\ &
  \ds
 +\frac{1}{|\S_1^d|}\int_0^t ds
  A(s,t)^{-d/2} M_0'(s)\int_{\sigma \in \S_1^d}
  d\sigma \ \exp\left(-\frac{R^2|B(s,t)\sigma +\omega|^2}{4A(s,t)}\right) =0.
\end{array}
\ee
Note that this last relation does not depend on $\omega$ and in particular can be averaged over $\omega \in \S_1^d$.
Finally  $M_0'(s)$ is the derivative of $M_0(s)$ at time $s$.
\end{proposition}

The proof of this result is given in Appendix \ref{sec_a1}. Note that
the solution
\fref{solution-exacte-d} is semi-explicit in the sense that it
involves the quantity $M_0(s)$ which  depends on the
solution itself. However, the only knowledge of the time evolution of
this mass allows the determination of the whole solution $f$. The exact
equation satisfied by {$M_0(s)$ is  given by  \fref{equation-masse} and, as it stands, seems to be too complicated
for practical use.
Nevertheless, this equation can be simplified in the
asymptotic limit $R\rightarrow +\infty$ for fixed time $t$. This is the subject of the following proposition

\vskip0.5cm

\begin{proposition}[Approximate mass law for large $R$]
\label{prop3}
Let $f$ be a smooth enough solution of \fref{FP} with $D=1$ and isotropic initial data $f_0$ supported on $B_R$. Let
$M_0(t)$ be  the total mass at time $t$ given by \fref{M0M1}. Then, for
any given time $t> 0$
\be
\label{M0Rgrand}
\begin{array}{ll} \ds
M_0'(t) \ \sim \ & \ds  - \frac{2^{\frac{2d-3}{2}}\beta^{3/2} |\S_1^d|}{\sqrt{\pi(1-\exp(-2 \beta
    t))}} R^{\frac{d+1}{2}}\exp\left(\frac{d-1}{2}\beta t\right)
      \\
    & \ds \times\int_0^{R} r^{\frac{d-1}{2}} f_0(r) \exp\left(-\frac{\beta
    \left(\exp(-\beta t)r- R\right)^2}{2(1-\exp(-2\beta t))}\right)
    dr,
\end{array}
\ee
as $R$ goes to $+\infty$.  Furthermore,

\begin{itemize}
\item If $f_0$ does not depend on $R$ and is supported on $[0,R_0]$
  with a non-vanishing left derivative $f_0'(R_0)$ at $R_0$, then
 \be
\label{f0fixe}
\begin{array}{ll} \ds
M_0'(t) \ \sim \   \frac{2^{\frac{2d+1}{2}} |\S_1^d|}{\sqrt{\pi\beta (1-\exp(-2 \beta
    t))}} & \ds R^{\frac{d-3}{2}}
    R_0^{\frac{d-1}{2}}\exp\left(\frac{d-1}{2}\beta t\right)
\sinh^{2}(\beta t)    \\ & \ds \times\exp\left(-\frac{\beta
    \left(\exp(-\beta t)R_0 - R\right)^2}{2(1-\exp(-2\beta t))}\right)
    f_0'(R_0),
\end{array}
\ee
as $R\to +\infty$.
\item If $f_0$ depends on $R$  with a non vanishing left  derivative $f_0'(R)$  at $R$,
  then
\be
\label{f0nonfixe}
\begin{array}{ll} \ds
M_0'(t) \ \sim \ &  \ds \frac{2^{\frac{2d+1}{2}}  |\S_1^d|}{\sqrt{\pi\beta (1-\exp(-2 \beta
    t))}}  R^{d-2}\exp\left(\frac{d-1}{2}\beta t\right)  \\  & \ds
 \times\frac{\sinh^{2}(\beta t)}{(1-\exp(-\beta t))^{2}}\exp\left(-\frac{\beta
    \left(1- \exp(-\beta t)\right)^{2}}{2(1-\exp(-2\beta t))} R^2\right)
    f_0'(R).
  \end{array}
\ee
as $R\to+ \infty$.
\end{itemize}
\end{proposition}
The proof of this result is given in Appendix \ref{sec_a2}.

{\em Remark:} In practice, for sufficiently large $R$, this expression
is `valid' for $t\ll R^2$ so that $M_0'(t)\ll 1$. This
clearly shows that the order of the limits $R\rightarrow +\infty$ and
$t\rightarrow +\infty$ is not interchangeable. Usually, most works
\cite{chandra,chandramore,sh,michie,king} consider the limit
$t\rightarrow +\infty$, then the limit $R\rightarrow +\infty$. By
contrast, the above expressions are valid for $R\rightarrow +\infty$
at any fixed time $t$.

\subsection{Exact rate of escape for \fref{FP}}

In this section, we focus on the Fokker-Planck equation \fref{FP} with
$D\equiv 1$.  It is well known that the long time behavior
$t\rightarrow +\infty$ of the solution to \fref{FP} can be described
from the knowledge of the first eigenvalue (the largest nonzero
eigenvalue) of the linear operator $Q^{FP}_R$ constrained with the
vanishing Dirichlet boundary condition. This is essentially a
consequence of the self-adjointness of this operator in the space
$L^2\left(\exp(|v|^2/2)dv\right)$. However, the analytical
determination of this eigenvalue is a difficult task.  To our
knowledge, the first work on this subject goes back to Chandrasekhar
\cite{chandra} but the expression obtained for the fundamental
eigenvalue is only an approximation in the limit of large $R$ and is
given in terms of a (not explicitly summable) series. Here, we propose
another strategy which is based on the semi-explicit solution
\fref{solution-exacte-d} and derive an exact autonomous
relation satisfied by this first eigenvalue for any sufficiently large
$R$. Then, as a consequence, we recover the Chandrasekhar result (in a
more explicit form) by taking the leading term in our relation when
$R$ is large.  Here is the statement

\begin{proposition}[Exact rate of escape for \fref{FP}]
\label{prop4}
Let $\lambda (R)$ be the largest non zero eigenvalue of the linear
operator $Q_{FP}$ given by \fref{FP} with $D\equiv 1$, in the space of
isotropic functions of  $L^2\left(\exp(|v|^2/2)dv\right)$,  vanishing at the boundary.  Then

\noindent i) $\lambda=\lambda(R)$ is negative and, for sufficiently large $R$ so that $\lambda(R)+2\beta>0$, it  satisfies the following nonlinear relation
\be
\label{relation-lambda}
G(R\omega,0) + \frac{\lambda}{\beta} \int_0^1 \left(G(R\omega,u)- G(R\omega,0)  \right)
u^{\frac{\lambda}{\beta} -1}du=0, \ \ \ \ \forall \ \omega \in \S_1^d,
\ee
where
\be
\label{G}
G(v,u)=
\frac{1}{(2\pi)^d} \frac{1}{|\S_1^d|}\left(\frac{2\pi\beta}{1-u^2}\right)^{d/2}\int_{\sigma \in \S_1^d}
  \exp\left(-\frac{\beta|R u\sigma +v|^2}{2(1-u^2)}\right)\, d\sigma,
\ee
for all $v\in \R^d$ and $u\in [0,1[$. Note that relation \fref{relation-lambda} is independent of $\omega$ and can therefore be averaged  over $\omega
\in \S_1^d$.

\noindent ii) The corresponding eigenfunction is exactly given by 
\be
\label{eigenfunction}
f_{\lambda}(v)=M_{\lambda}\left\lbrace G(v,0)+\frac{\lambda}{\beta}\int_{0}^{1}\left\lbrack G(v,u)-G(v,0)\right \rbrack u^{\frac{\lambda}{\beta}-1}\, du\right\rbrace,
\ee
with $M_{\lambda}=\int_{B_R} f_{\lambda}(v)dv$.

\noindent iii) The eigenvalue $\lambda(R)$ has the following asymptotic
behavior
\be
\label{lambda-asymptotique}
\lambda (R) \ \sim \ - \ \frac{2\beta}{\Gamma \left(\frac{d}{2}\right)} \left(\frac{\beta
 R^2}{2}\right)^{d/2} \exp\left(\frac{-\beta R^2}{2}\right),
\ee
as $R$ goes to $+\infty$.
\end{proposition}
The proof of this result is given in Appendix \ref{sec_a3}.

{\it Remark 1}: The explicit asymptotic behavior
\fref{lambda-asymptotique} was not given by Chandrasekhar
\cite{chandra} who obtained the asymptotic expression of the
eigenvalue in the form of a series. In Appendix
\ref{sec_comp}, we obtain the asymptotic behavior
\fref{lambda-asymptotique} by a different method which can be extended
to the case of a Fokker-Planck equation with an arbitrary diffusion
coefficient $D(|v|)$ and potential $U(|v|)$.} However, the method in
Appendix \ref{sec_comp} is formal and, as it stands, cannot be
considered as complete mathematical proof of the result, unlike the proof given in Appendix \ref{sec_a3}.

{\it Remark 2}: The expression of the function $G(v,u)$ can be simplified as shown in Appendix \ref{sec_Gsimple}.

\section{Numerical results and discussion}
\label{sec_figures}

We have performed numerical simulations in order to illustrate our
theoretical results. We have taken $d=3$ (appropriate to stellar
systems) and $D=\beta=1$.

\begin{figure}[htbp]
\centerline{
\includegraphics[width=7cm,angle=0]{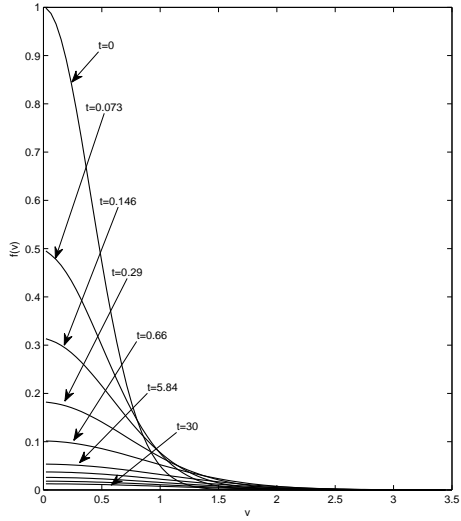}
} \caption[]{Velocity distribution $f(v,t)$ at different times. The parameters are $d=3$, $D=1$, $\beta=1$, $R=2\sqrt{3}$. The initial distribution is $f_0(v)=e^{-3v^2}$ for $v\le 1$ and $f_0(v)=0$ for $v>1$. Due to evaporation, the distribution function decreases and tends to zero for $t\rightarrow +\infty$. } \label{distri}
\end{figure}

\begin{figure}[htbp]
\centerline{
\includegraphics[width=8cm,angle=0]{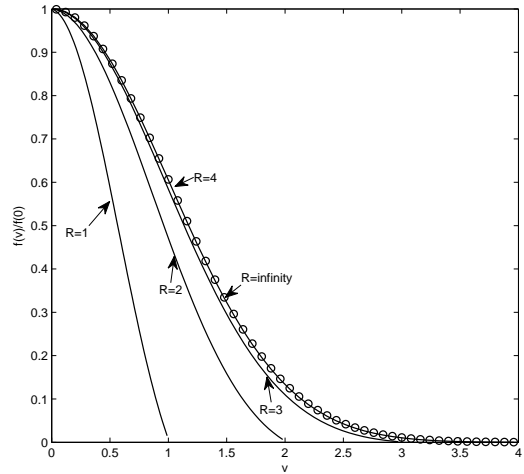}
} \caption[]{Normalized velocity profile  $f(v)/f(0)$ at large times (eigenfunction) for different values of $R$. The slope at the escape velocity $v=R$ decreases as $R$ increases leading to slower mass loss. For $R\rightarrow +\infty$, the distribution function tends to the Maxwellian (bullets). } \label{fpro1}
\end{figure}

\begin{figure}[htbp]
\centerline{
\includegraphics[width=8cm,angle=0]{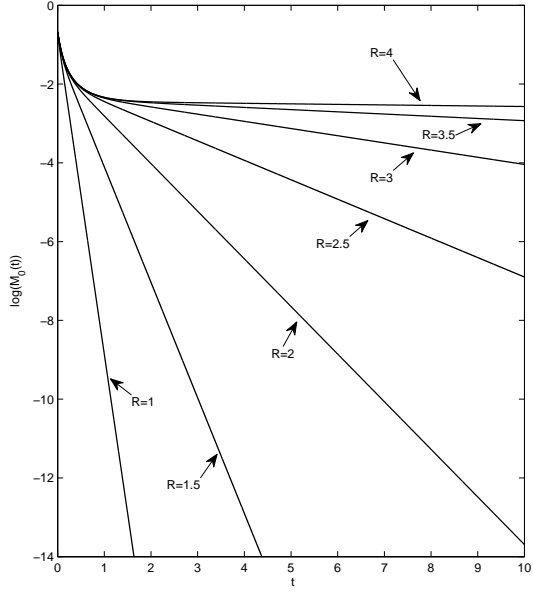}
} \caption[]{Decay of the mass $M_0(t)$ contained in the cluster as a function of time (logarithmic scale). For large times, the decay is exponential $M_0(t)\sim e^{-|\lambda(R)|t}$ leading to straight lines with slope $\lambda(R)$. The exponential decay rate $|\lambda(R)|$ decreases as $R$ increases.  } \label{mass}
\end{figure}

%\begin{figure}[htbp]
%\centerline{
%\includegraphics[width=8cm,angle=0]{fpro2.eps}
%} \caption[]{The function $\gamma(t)$.} \label{gamma}
%\end{figure}

\begin{figure}[htbp]
\centerline{
\includegraphics[width=8cm,angle=0]{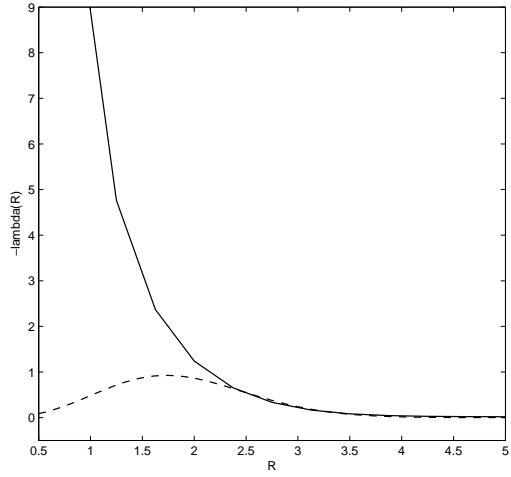}
} \caption[]{Mass decay rate $|\lambda(R)|$ (eigenvalue) for different
values of $R$. The decay rate tends to $+\infty$ for $R\rightarrow 0$
and is equivalent to the asymptotic expression
(\ref{lambda-asymptotique}) for $R\rightarrow +\infty$.  } 
\label{lambda}
\end{figure}

In Fig. \ref{distri}, we show the velocity distribution $f(v,t)$ at
different times. We have adopted the value $R=2\sqrt{3}$ of the escape
velocity corresponding to the estimate deduced from the virial theorem (see
Introduction). The initial distribution is $f_0(v)=e^{-3v^2}$ for
$v\le 1$ and $f_0(v)=0$ for $v>1$. Due to evaporation, the
distribution function decreases and tends to zero for $t\rightarrow
+\infty$. In fact, its large time behavior is of the form $f(v,t)\sim
e^{-|\lambda(R)|t}f_{\lambda}(v)$. If we rescale the distribution
$f(v,t)$ by its central value $f(0,t)$, then the normalized velocity
profile $f(v,t)/f(0,t)$ tends to the eigenfunction
$f_{\lambda}(v)/f_{\lambda}(0)$. This eigenfunction is represented in
Fig. \ref{fpro1} for different values of $R$. We see that the slope of the
distribution at the escape velocity $R$ decreases as $R$ increases
implying a slower mass loss. In fact, for $R\rightarrow +\infty$, the
mass loss tends to zero and the eigenfunction tends to the Maxwellian
which is the steady state of the Kramers equation without velocity
confinement. In Fig. \ref{mass}, we plot the mass $M_0(t)$ contained in the
cluster as a function of time in logarithmic scale. For large times,
the decay is exponential $M_0(t)\sim e^{-|\lambda(R)|t}$ leading to
straight lines with slope $\lambda(R)$. The exponential decay rate
$|\lambda(R)|$ decreases as $R$ increases. This decay rate is plotted
in Fig. \ref{lambda} for different values of $R$. The decay rate tends to
$+\infty$ for $R\rightarrow 0$ and behaves like (\ref{lambda-asymptotique}) for $R\rightarrow +\infty$.

\section{Conclusion}
\label{sec_conclusion}

In this paper, we have obtained new analytical results for the escape
of stars from globular clusters in the framework of the Chandrasekhar
model. These results can also have applications to other physical
systems described by the Kramers equation with parabolic potential and
absorbing boundary conditions (i.e. the classical Kramers problem).

We have first obtained an autonomous equation for the mass loss [see
\fref{equation-masse}] and a semi-explicit expression of the
distribution function $f(v,t)$ [see \fref{solution-exacte-d}] that
are valid for an arbitrary escape velocity $R$ and time $t$.  We have
simplified the expression of the mass loss in the limit $R\rightarrow
+\infty$ for any fixed time $t$ [see \fref{M0Rgrand}]. We have also
used the semi-explicit expression of the distribution function
$f(v,t)$ to obtain an exact integral equation for the fundamental
eigenvalue $\lambda(R)$ [see \fref{relation-lambda}] and for the
fundamental eigenfunction $f_{\lambda}(v)$ [see \fref{eigenfunction}]
that are valid for any sufficiently large $R$. This is an interesting
complement to the perturbative results derived by Chandrasekhar
\cite{chandra} that are valid in the asymptotic limit $R\rightarrow
+\infty$. We have obtained the explicit behavior of the fundamental
eigenvalue in the limit $R\rightarrow +\infty$ [see
\fref{lambda-asymptotique}] which improves upon the result of
Chandrasekhar expressed in the form of a series. Additional asymptotic
results for the fundamental eigenvalue and fundamental eigenfunction
are given in Appendix \ref{sec_comp}. Finally, we have illustrated our
results with numerical simulations [see Sec. \ref{sec_figures}].

Of course, our approach is based on several approximations. As
previously discussed, it assumes that the medium is infinite and
spatially homogeneous and that the encounters between stars can be
described by the Kramers equation (Chandrasekhar's model). Furthermore,
our analytical results (except those of Appendix \ref{sec_comp}) are
valid only when the diffusion coefficient in the Fokker-Planck
equation is constant while a more exact description of encounters
between stars would involve a velocity dependent diffusion coefficient
\cite{chandramore,michie,king}. 
It remains therefore a challenging issue to extend the mathematical
theory of the escape of stars from globular clusters in the case of
more realistic models.

\vskip0.2cm

\begin{acknowledgement} M. Lemou was supported by the French  Agence Nationale de la Recherche, ANR JC MNEC.
\end{acknowledgement}

\appendix

\section{Proof of Proposition \ref{prop2}}
\label{sec_a1}

We consider the  Fokker-Planck equation
\fref{FP} with $D=1$.  We recall that the initial data is assumed to be isotropic
(i.e. it depends on the modulus of the velocity only), which implies
that the solution $f(v,t)$ is isotropic at any time.
% The strategy has some similarities with the one used in  \cite{Lemou1}.

First we claim that if $f$ is a solution  to \fref{FP} on the ball $B_R$ with a
vanishing boundary condition, then $f$ is a solution on all $\R^d$ in
a distributional sense to
\be
\label{FP-Dirac}
 \frac{\partial f}{\partial t}  (v,t) =\nabla \cdot
         \left(  \nabla f(v) + \beta f(v) v\right) + M_0'(t)\frac{\delta_{\S_R^d}}{|\S_R^d|},
\ee
where $M_0$ is  related to $f$  by \fref{M0M1} and $\delta_{\S_R^d}$ is
         the Dirac operator on the sphere:
$$
  <\delta_{\S_R^d} , \phi>
 =\int_{\S_R^d} \phi(v) d\sigma(v),$$
for all test function $\phi$. Note that on isotropic test functions
  $\phi$, this Dirac operator reduces to: $ <\delta_{\S_R^d} , \phi> = |\S_R^d|\phi(R)$.

% Indeed, let
%$$M_n(t)= \int_{B_R} |v|^{2n} f(t,v)dv, \ \ \ n\in \N$$
%and inetgrate \fref{FP} against $|v|^{2n}, n\geq 1$. After easy computation,
%one gets
%$$M_n'(t)= R^{2n}  \int_{\S_R} \nabla f \cdot n d\sigma - 2\beta n
%M_n(t) + 2n (2n+d-2) M_{n-1)(t).$$
%Observing that if we do the same for $n=0$, we get
%$$M_0'(t)= \int_{\S_R} \nabla f \cdot n d\sigma,$$
%and then
%
%$$M_n'(t)= R^{2n} M_0'(t) - 2\beta n
%M_n(t) + 2n (2n+d-2) M_{n-1)(t), \ \ \ \ n\geq 1 $$
To prove \fref{FP-Dirac}, consider an isotropic solution $f$ to
\fref{FP} which we extend to $0$ outside $B_R$, and integrate
\fref{FP} against an isotropic test function $\phi$ on $\R^d$.  We
obtain after integration by parts:
$$ \frac{d}{dt} \left(\int_{\R^d} f(v,t)\phi(|v|)dv\right)= \int_{\R^d}
\left( \Delta \phi - {\beta} v\cdot \nabla\phi\right) f(v) dv   + \left(
\int_{\S_R^d} \nabla f\cdot n \ d\sigma(v)\right) \phi(R).$$
Taking $\phi\equiv 1$ on $B_R$, we also have
$$M_0'(t)= \int_{\S_R^d} \nabla f \cdot n \ d\sigma(v).$$
Finally, these last two identities imply that $f$ is a solution to
\fref{FP-Dirac} in the sense of distributions.  Note that the choice
 of isotropic test functions $\phi$ is sufficient thanks to the radial
 symmetry of $f$.

{\it Remark}: We have shown that the solution of \fref{FP} is solution of
\fref{FP-Dirac}. This is sufficient for our purposes. However, it is
easy to prove that, reciprocally, \fref{FP-Dirac} admits an infinity
of smooth solutions and that, among all of them, the solution that
satisfies $f(R\omega,t)=0$ is the solution of \fref{FP}.

%we now consider the radial solution $g$ to the following linear equation on $\R^d$
%\be
%\label{FP-g-Delta}\frac{\partial g}{\partial t}  (t,v) =\nabla \cdot
%         \left(\nabla g(v) + \beta v g(v)\right) +
%         M_'(t)\f0rac{\delta_{\S_R^d}}{|\S_R^d|},
%\ee
%with the initial condition
%$g(0,v)= f_0(v)$ for $|v|\leq R$, and $g(0,v)=0$ elswhere.
%We integrate \fref{FP-g-Delta} with respect to $v$ and obtain
%$$(M_0^g)'(t) = M_0'(t) \ \ \mbox{with} \ \
%M_0^g(t) =\int_{R^d} g(t,v) dv. $$
%Therefore $f$ and $g$ satisfy the  same differential equation and $g(0,v)= f_0(v)$.  We
%conclude that
%$$g(t,v) = f(t,v),  \ \ \forall \ t\leq 0, \ \ v\in\R^d,$$
%$f$ being extended to $0$ outside $B_R$.

We shall now use Fourier transform techniques to find out a
semi-explicit solution to \fref{FP-Dirac}.  Let $\fc (\xi,t)$ be the
Fourier transform of $f$ in the $v$ variable: $$\fc (\xi,t)=
\int_{\R^d} f(v,t) \exp(-iv\cdot \xi) \, dv,$$ $$f (v,t)= \int_{\R^d}
\fc(\xi,t) \exp(iv\cdot \xi) \, \frac{d\xi}{(2\pi)^d}.$$ Taking the
Fourier transform in (\ref{FP-Dirac}), we get $$\frac{\partial
\fc}{\partial t} (\xi,t)= -\xi^2\fc - \beta \xi\cdot\nabla\fc +
M_0'(t) H(\xi),$$ where $$H(\xi)= \frac{1}{|\S_R^d|} \int_{\S_R^d}
\exp(-iv\cdot \xi)\, d\sigma({v}).$$ We now use the method of
characteristics to solve this equation.  Let $$F(\xi,t) = \fc(\xi
\exp(\beta t),t).$$ Easy computations yield $$\frac{\partial
F}{\partial t} (\xi,t)= -\xi^2\exp(2\beta t) F(\xi,t) + M_0'(t) H(\xi
\exp(\beta t)).$$
This can be written as
$$\frac{\partial}{\partial t} \left(
 \exp\left(\frac{\xi^2}{2\beta}(\exp(2\beta t)-1)\right) F(\xi,t)\right)=
 \exp\left(\frac{\xi^2}{2\beta}
(\exp(2\beta t)-1)\right)M_0'(t) H(\xi \exp(\beta t)).$$
Integrating over time, we get
$$\begin{array}{ll} \ds F(\xi,t)= &
  \exp\left(-\frac{\xi^2}{2\beta}(\exp(2\beta t)-1)\right) F(\xi,0)\\
  & \ds
+ \int_0^t \exp\left(-\frac{\xi^2}{2\beta}(\exp(2\beta t)-\exp(2\beta
s))\right)M_0'(s) H(\xi \exp(\beta s))ds.
\end{array}$$
 As $\fc(\xi,t) = F(\xi \exp(-\beta t),t)$, we obtain
$$\begin{array}{ll}\ds \fc(\xi,t)= & \ds \exp\left(-\frac{\xi^2}{2\beta}(1-\exp(-2\beta t))\right)
 \fc(\xi\exp(-\beta t),0) \\
& \ds  + \int_0^t \exp\left(-\frac{\xi^2}{2\beta}(1-\exp({-}2\beta(t-
s))\right)M_0'(s) H(\xi \exp(-\beta (t-s)))\, ds.
\end{array}$$
Using the notations of proposition \ref{prop2}, this can also be
 written as
$$\fc(\xi,t)= \exp\left(-\xi ^2 A(0,t)\right)
 \fc(\xi B(0,t),0)
+ \int_0^t \exp\left(-\xi^2 A(s,t)\right)M_0'(s) H(\xi B(s,t))\, ds.$$
Now, we take the inverse Fourier transform using the identities
$${\rm Inv}F\left(\exp(-A\xi^2)\right)(v) = \frac{1}{(2\pi)^d}
\left(\frac{\pi}{A}\right)^{d/2} \exp\left(
-\frac{|v|^2}{4A}\right),$$
and
$${\rm Inv}F\left({\exp(-A\xi^2) \exp\left(-iB\xi\cdot w\right)}\right)(v) = \frac{1}{(2\pi)^d}
\left(\frac{\pi}{A}\right)^{d/2} \exp\left(
-\frac{|Bw-v|^2}{4A}\right),$$
to  get the desired expression \fref{solution-exacte-d}.

{\it Remark:} if we integrate \fref{solution-exacte-d} on the
velocity, we find a trivial result: $M_{0}(t)=M_{0}(t)$, which shows
the consistency of this equation.

%%%%%%%%%%%%%%%%%%%%%%%%%%%%%%%%%%%%%
\section{Proof of Proposition \ref{prop3}}
\label{sec_a2}

Let $f$ be the solution to \fref{FP} with initial data $f_0$ and let
$M_0$ be defined by \fref{M0M1}. From \fref{solution-exacte-d} and the
vanishing boundary condition $f(R\omega,t)=0,$ $\omega \in {S_1^d},$ we
get
\be
\label{T1T2=0}
T_1+T_2=0,
\ee
where
\be
\label{T1}
T_1=|\S_1^d| A(0,t)^{-d/2} \int_{\R^d} f_0(v_*)\,
  \exp\left(-\frac{|R\omega-B(0,t)v_*|^2}{4A(0,t)}\right) \, dv_*,
\ee
\be
\label{T2}
T_2=\int_0^t ds A(s,t)^{-d/2} M_0'(s)\int_{\sigma \in \S_1^d}
   \exp\left(-\frac{R^2|B(s,t)\sigma +\omega|^2}{4A(s,t)}\right)  \, {d\sigma},
\ee
where $A$ and $B$ are defined in proposition \ref{prop2}.

To analyze  the asymptotic behavior of the  term $T_1$ when $R$
  goes to $+\infty$, the time $t$ being fixed, we introduce a spherical system of coordinates (in $\R^d$), write $v_*=r\sigma$, $\sigma
  \in \S_1^d$,  $\sigma\cdot\omega=\cos\theta$ and obtain
$$
\begin{array}{ll}
\ds T_1& \ds=|\S_1^d| A(0,t)^{-d/2} \int_{0}^{+\infty}\int_{0}^{\pi} f_0(r)\,\\
  \ds& \ds\times\exp\left(-\frac{R^2-2RB(0,t)r\cos\theta+B(0,t)^2r^2}{4A(0,t)}\right) \, C_d r^{d-1}dr (\sin\theta)^{d-2}d\theta,
\end{array}
$$
where  $C_d$ is given by
\be
\label{Cd}C_d= \frac{|\S_1^d|}{\ds \int_0^\pi (\sin\theta)^{d-2} d\theta},  \  \  \   d\geq 2.
\ee
In the following calculations, we shall assume $d\ge 2$ but we have
checked by a specific calculation that the final result remains valid
for $d=1$. Making the change of variable $\delta=-\cos\theta$, we obtain
$$
\begin{array}{ll}
 \ds T_1= & \ds C_d |\S_1^d|  A(0,t)^{-d/2}   \exp\left(-\frac{R^2}{4A(0,t)}\right)  \\   &
   \ds  \times\int_0^{+\infty} \int_{-1}^{1}
   f_0(r) \, \exp\left(-\frac{B(0,t)^2}{4A(0,t)} r^2  \right)
\exp\left(-\frac{B(0,t)}{2A(0,t)} R r \delta  \right) r^{d-1} \left(1- \delta^2 \right)^{\frac{d-3}{2}} \, d\delta dr.
\end{array}
$$
We then make the change of variable   $u= \frac{B(0,t)}{2A(0,t)}R r(1+\delta)$ in the integral
  over $\delta$, and get
$$\begin{array}{ll}
\ds & \ds T_1= C_d 2^{d-2}|\S_1^d|
 A(0,t)^{-d/2} \left(\frac{A(0,t)}{B(0,t)R}\right)^{\frac{d-1}{2}}
   \\ & \ds
\times \int_0^{+\infty}
  \int_{0}^{\frac{B(0,t)}{A(0,t)}R r} f_0(r)\, \exp\left(-\frac{(B(0,t)r-R)^2}{4A(0,t)}
  \right) \left( 1- \frac{A(0,t)}{B(0,t) R}\frac{u}{r}
  \right)^{\frac{d-3}{2}}   r^{\frac{d-1}{2}} u ^{\frac{d-3}{2}} \exp(-u) \, du dr.
\end{array}$$
Now let $R$ go to infinity to obtain
\be
\label{equiv-T1}
\begin{array}{ll} \ds
T_1\ \sim \  C_d 2^{{d-2}}|\S_1^d| & \ds
\Gamma\left(\frac{d-1}{2}\right)  A(0,t)^{-d/2} \left(\frac{A(0,t)}{B(0,t)R}\right)
^{\frac{d-1}{2}}  \\ & \ds\times \int_0^{+\infty}
  f_0(r)\, \exp\left(-\frac{(B(0,t)r-R)^2}{4A(0,t)}
  \right)r^{\frac{d-1}{2}}\, dr,
\end{array}
\ee
as $R$ goes to $+\infty$.

Assume first that $f_0$ does not depend on $R$ and is supported on
$[0,R_0]$.  In that case, we perform the change of variable:

$$ s= \frac{B(0,t)R}{2A(0,t)} (R_0-{r}),$$ and get

$$ \begin{array}{ll}  \ds &\ds\int_0^{+\infty}
   \ds f_0(r)\, \exp\left(-\frac{(B(0,t)r-R)^2}{4A(0,t)}
  \right)r^{\frac{d-1}{2}} \, dr \ =   \ \frac{2A(0,t)}{B(0,t)R}\, \exp\left( -
\frac{(B(0,t)R_0-R)^2}{4A(0,t)}\right) \\ &  \ds  \times\int_0^{\frac{B(0,t)R
 R_0}{2A(0,t)}}
 f_0\left(
 R_0 -\frac{2A(0,t)}{B(0,t)R}s \right) \exp \left( -s +
 \frac{B(0,t)R_0}{R} s - \frac{A(0,t)}{R^2}s^2\right)\left( R_0 -\frac{2A(0,t)}{B(0,t)R}s \right)^{\frac{d-1}{2}}\, ds.
\end{array}
$$
If we let $R$ go to infinity in this expression, then we obtain
\be
\label{eq1}
\begin{array}{ll}\ds
\int_0^{+\infty}
  f_0(r) & \ds  \exp\left(-\frac{(B(0,t)r-R)^2}{4A(0,t)}
  \right)r^{\frac{d-1}{2}}\, dr \ \sim \\ & \ds \ -\left(\frac{2A(0,t)}{B(0,t)R}\right)^2 \exp\left( -
\frac{(B(0,t)R_0-R)^2}{4A(0,t)}\right) R_0^{\frac{d-1}{2}} f_0'(R_0),
\end{array}
\ee
for large  $R$.
The case where $f_0$ does depend on $R$ can be treated similarly to
yield
\be
\label{eq2}
\begin{array}{ll} \ds
\int_0^{+\infty}
  f_0(r) & \ds \exp\left(-\frac{(B(0,t)r-R)^2}{4A(0,t)}
  \right)r^{\frac{d-1}{2}}\, dr \ \sim \\  & \ds  -\left(\frac{2A(0,t)}{B(0,t)R}\right)^2
 \exp\left( -\frac{(1-B(0,t))^2}{4A(0,t)} R^2\right)
 \frac{R^{\frac{d-1}{2}}}{\lbrack 1-B(0,t)\rbrack^{2}} f_0'(R),
\end{array}
\ee
for large $R$.  We now deal with the asymptotic behavior of the term
$T_2$ given by
\fref{T2} when $R$ goes to $+\infty$. First, we write $T_2$ in the
following form
$$
T_2=\int_0^t ds A(s,t)^{-d/2} M_0'(s)\exp\left(-\frac{R^2(B(s,t)^2 +1)}{4A(s,t)}\right) \int_{\sigma \in \S_1^d}
   \exp\left(-\frac{R^2B(s,t)}{2A(s,t)}\sigma \cdot \omega\right)\, d\sigma.
$$
We then introduce spherical coordinates (in $\R^d$) in the variable
$\sigma \in \S_1^d$ and obtain
$$
T_2=C_d\int_0^t ds A(s,t)^{-d/2} M_0'(s)\exp\left(-\frac{R^2(B(s,t)^2 +1)}{4A(s,t)}\right) \int_{0}^\pi
   \exp\left(-\frac{R^2B(s,t)}{2A(s,t)}\cos\theta\right) (\sin \theta)^{d-2}\, d\theta.
$$
 Performing  the change of variable $u =
 \frac{R^2B(s,t)}{2A(s,t)}(1+\cos \theta)$ in the integral over $\theta$,
we get
$$
\begin{array}{ll} \ds
T_2=C_d  2^{{d-2}}\int_0^t ds A(s,t)^{-d/2} M_0'(s) & \ds
\left(\frac{A(s,t)}{R^2 B(s,t)}\right)^{\frac{d-1}{2}}
\exp\left(-\frac{R^2(B(s,t)-1)^2}{4A(s,t)}\right)\\
& \ds  \times \int_{0}^{\frac{R^2 B(s,t)}{A(s,t)}}
  \left(1- \frac{A(s,t)}{R^2 B(s,t)} u\right)^{\frac{d-3}{2}}
  u^{\frac{d-3}{2}} \exp(-u) \, du.
\end{array}
$$
Letting $R\rightarrow +\infty$, we have
$$
T_2\ \sim \ C_d  2^{d-2}\Gamma\left(\frac{d-1}{2}\right)\int_0^t  A(s,t)^{-d/2} M_0'(s) \left(\frac{A(s,t)}{R^2 B(s,t)}\right)^{\frac{d-1}{2}}\exp\left(-\frac{R^2(B(s,t)-1)^2}{4A(s,t)}\right)\, ds.
$$
We now use  the following  change of variable in this last integral
over $s$:
$$ \tau = \frac{1-B(s,t)}{2\sqrt{A(s,t)}} R= \left(\frac{\beta}{2}\frac{1-
    \exp(-\beta(t-s))}{1+ \exp(-\beta(t-s))}\right)^{1/2} R,
$$ to obtain
%$$
%\begin{array}{ll} \ds
%\textcolor{green}{T_2\ \sim \ C_d
%2^{\frac{d-3}{2}}} & \ds \textcolor{green}{\Gamma\left(\frac{d-1}{2}\right)\frac{8}{\beta\sqrt{2}}
%R^{-d} \int_0^{R\left( \frac{\beta(1-\exp(-\beta t))}{2(1+\exp(-\beta
%    t))}\right)^{1/2}}
%\left( 1-\frac{2\tau^2}{\beta R^2}\right)^{d/2}  \left(
%1-\frac{2\tau^2}{\beta R^2}\right)^{\frac{1-d}{2}}} \\ & \ds
%\textcolor{green}{\exp(-\tau^2)  M_0'\left(t+ \frac{1}{\beta} \ln \left(\frac{\beta R^2 -
%  2\tau^2}{\beta R^2 + 2\tau^2} \right)\right) d\tau.}
%\end{array}
%$$
$$
\begin{array}{ll} \ds
T_2\ \sim \ C_d
2^{d-2} & \ds \Gamma\left(\frac{d-1}{2}\right)\frac{4}{\beta}
R^{-d} \int_0^{R\left( \frac{\beta(1-\exp(-\beta t))}{2(1+\exp(-\beta
    t))}\right)^{1/2}}
\left( 1-\frac{2\tau^2}{\beta R^2}\right)^{-(d+1)/2}  \left(
1+\frac{2\tau^2}{\beta R^2}\right)^{\frac{d-1}{2}} \\ & \ds
  M_0'\left(t+ \frac{1}{\beta} \ln \left(\frac{\beta R^2 -
  2\tau^2}{\beta R^2 + 2\tau^2} \right)\right) \exp(-\tau^2)\, d\tau.
\end{array}
$$
This yields an equivalent of $T_2$
%\be
%\label{equiv-T2}
%\textcolor{green}{T_2\ \sim \ C_d
%2^{\frac{d-3}{2}}\Gamma\left(\frac{d-1}{2}\right)\frac{2\sqrt{2\pi}}{\beta}R^{-d} M_0'(t), }
%\ee
\be
\label{equiv-T2}
T_2\ \sim \ C_d
2^{{d-1}}\Gamma\left(\frac{d-1}{2}\right)\frac{\sqrt{\pi}}{\beta R^{d}} M_0'(t),
\ee
for large $R$.

Combining  \fref{T1T2=0} with \fref{equiv-T1} and
\fref{equiv-T2} leads to the result \fref{M0Rgrand} of proposition
\ref{prop3}.
Finally, substituting \fref{eq1} (resp. \fref{eq2}) into \fref{M0Rgrand}
yields \fref{f0fixe} (resp. \fref{f0nonfixe}).

%%%%%%%%%%%%%%%%%%%%%%%%%%%%%%%%%%%%%%%%
\section{Proof of proposition \ref{prop4}}
\label{sec_a3}

We first prove i) in proposition \ref{prop4}. We shall use the exact solution
\fref{solution-exacte-d} stated in  proposition \ref{prop2}. Let
$\lambda$ be the  fundamental ({largest} non zero) eigenvalue of the Fokker-Planck operator
\fref{FP}. Even if the proof of $\lambda <0$  is classical,  we
give the argument here for the sake of completeness. If $f$ is  an associated
eigenfunction, the eigenvalue problem can be written as
$$\nabla\cdot \left[
         D(|v|)\exp(-\beta |v|^2/2) \nabla \left(f(v)\exp(\beta |v|^2/2)
         \right) \right] =   \lambda f.$$
Integrating  against $f(v)\exp(\beta |v|^2/2)$ on $B_R$, we get
$$-\int_{B_R}  D(|v|)\exp(-\beta |v|^2/2) \left|\nabla
\left(f(v)\exp(\beta |v|^2/2)\right)\right|^2 dv = \lambda
\int_{B_R}f(v)^2\exp(\beta |v|^2/2)dv,$$
which implies {that} $\lambda <0$.  We now prove relation
\fref{relation-lambda}.  Let $f_\lambda(v)$ be an
{eigenfunction} associated to $\lambda$, then
$\exp(\lambda t) f_\lambda(v)$ is the solution of \fref{FP} with
initial data $f_{\lambda}$. Therefore, using \fref{solution-exacte-d},
the eigenvalue problem can be written as

$$\exp(\lambda t) f_\lambda(v) = T_1(f_\lambda)(v,t) + \lambda M_\lambda \int_0^t
    G(v,\exp(-\beta(t-s))\exp(\lambda s)\, ds,$$
where $G$ is given by \fref{G}, $M_\lambda = \int_{B_R}f_\lambda(v)
    dv$ and
\be
\label{T1w}
T_1(f_\lambda)(v,t)= \frac{1}{(2\pi)^d}
  \left(\frac{\pi}{A(0,t)}\right)^{d/2} \int_{{B_R}} f_\lambda(v_*)
  \exp\left(-\frac{|v-v_* B(0,t)|^2}{4A(0,t)}\right) dv_* .
\ee
This relation can be rewritten equivalently
\be
\label{eq3}
\begin{array}{ll}\ds \exp(\lambda t) f_\lambda(v) & \ds =T_1(f_\lambda)(v,t) - 
    M_{\lambda}G(v,0)+M_{\lambda}G(v,0)\exp(\lambda t)
    \\ & \ds {+}\lambda M_\lambda \int_0^t \left[G(v,\exp(-\beta(t-s))- G(v,0)\right]\exp(\lambda s)ds, \ \ \forall t\geq
    0.
\end{array}
\ee

We now {analyze} this relation for large time $t$. First, we expand the
term $ T_1(f_\lambda)(v,t)$ and obtain after easy computations:

$$\begin{array}{ll}\ds & T_1(f_\lambda)(v,t)=  \ds
   {G(v,0)\left[ M_\lambda +  \beta v\cdot 
  \left(\int  v_* f_\lambda(v_*) dv_*\right)\exp(-\beta t) +\frac{1}{2}(d-\beta v^2)M_\lambda\exp(-2\beta t)\right.}\\ & {\left.\ds
  -\frac{1}{2}\beta \left(\int |v_*|^2 f_\lambda(v_*)dv_*\right) \exp(-2\beta t)+\frac{1}{2}\beta^2  \left(\int (v \cdot v_*)^2 f_\lambda(v_*)dv_*\right)\exp(-2\beta t)\right]+
  O(\exp(-3\beta t)).}
\end{array}$$
As $f_{\lambda}$ is supposed to be radially symmetric in $v$, the  order 1 contribution in $\exp(-\beta t)$ vanishes, and we get after some rearrangements

$$
\begin{array}{ll}\ds T_1(f_\lambda)(v,t)=  \ds
  G(v,0)\left[ M_\lambda +\frac{1}{2}(d-\beta v^2)\left(M_\lambda  - \frac{\beta}{d}\int |v_*|^2 f_\lambda(v_*)dv_* \right)\exp(-2\beta t)\right] \\ \ds
   +
  O(\exp(-3\beta t)).
\end{array}
$$
This implies in particular that
$$\lim_{t\rightarrow + \infty} \exp(2\beta t)\left[
  T_1(f_\lambda)(v,t)- M_{\lambda}G(v,0)\right]$$
is finite.
Now we multiply  \fref{eq3}  by $\exp(2\beta t)$ and  perform
the change of variable $u=\exp(-\beta(t-s))$ in the rhs of \fref{eq3} to
obtain
\be
\label{eq4}
\begin{array}{ll}\ds
\exp\left( (\lambda
    +2\beta) t\right) f_{\lambda}(v)& =\exp(2\beta t)\left[T_1(f_\lambda)(v,t)-\right.  \left.
    M_{\lambda}G(v,0)\right]+ \ds M_{\lambda}\exp\left( (\lambda
    +2\beta) t\right) \\ & \ds \times \left[G(v,0) + \frac{\lambda}{\beta}\int_{\exp(-\beta t)}^1
    \left(G(v,u)- G(v,0)\right)u^{\frac{\lambda}{\beta}
    -1}du\right].
\end{array}
\ee
Before taking the limit $t\rightarrow +\infty$, we first claim that
$$\lambda + 2\beta >0,$$ at least for large enough $R$. Indeed, we
know that if $R=+\infty$, $0$ is an eigenvalue of the Fokker-Planck
operator \fref{FP}. Therefore, the first eigenvalue $\lambda(R)$ on
$B_R$ must go to $0$ when $R$ goes to infinity, and this proves the
claim.  Passing to the limit $t\rightarrow +\infty$ in
\fref{eq4}, we conclude that we necessarily have

$$f_{\lambda}(v)=M_{\lambda}\left\lbrace G(v,0) + \frac{\lambda}{\beta}\int_0^1
    \left[G(v,u)- G(v,0)\right]u^{\frac{\lambda}{\beta}
    -1}du \right\rbrace,$$
which is relation \fref{eigenfunction}. Finally, writing this relation at the {boundary} $v=R\omega$, $\omega\in \S_1^d$, and
 recalling that $f_\lambda(R\omega)=0$, we
  get

$$G(R\omega,0) + \frac{\lambda}{\beta}\int_0^1
    \left[G(R\omega,u)- G(R\omega,0)\right]u^{\frac{\lambda}{\beta}
    -1}du =0,$$
which is relation \fref{relation-lambda}.

We now prove  the asymptotic behavior \fref{lambda-asymptotique} for large
$R$. As $\lambda (R)$ goes to $0$ when $R$ goes to infinity and
$$ G(R\omega,0) = \frac{1}{(2\pi)^d} ({2\pi\beta})^{d/2} \exp\left( -
\beta \frac{R^2}{2}\right),$$
we obtain
$$
\begin{array}{ll}\ds \exp\left( -\beta \frac{R^2}{2}\right) \sim
     - \frac{\lambda}{\beta} \frac{1}{|\S_1^d|} \int_{\S_1^d} d\sigma \int_0^1\left[(1-u^2)^{-d/2} \exp\left( -\frac{\beta R^2}{2} \frac{|\sigma u
	+\omega|^2}{1-u^2}\right)- \exp\left( -\beta
	\frac{R^2}{2}\right)\right]\frac{du}{u}.
\end{array}$$
This is also equivalent to
\be
\label{eq5}
\lambda(R) \sim - \frac{\beta |\S_1^d|}{I_d(R)},
\ee
where
\be
\label{IdR}
I_d(R)=  \int_{\S_1^d} d\sigma \int_0^1
    \left[(1-u^2)^{-d/2} \exp\left( -\beta R^2\frac{u( u
	+\sigma\cdot \omega)}{1-u^2}\right)- 1\right]\frac{du}{u}.
\ee
Now we write
$$
I_d(R)= I_d^+(R)+  I_d^-(R),
$$
with
$$
I_d^+(R)=  \int_{\S^+} d\sigma \int_0^1
    \left[(1-u^2)^{-d/2} \exp\left( -\beta R^2\frac{u( u
	+\sigma\cdot \omega)}{1-u^2}\right)-
    1\right]\frac{du}{u}, $$
and $I_d^-(R)=I_d(R)-I_d^+(R)$, $\S^{+} =\{ \sigma \in \S_{1}^{d},  \sigma \cdot \omega > 0\}$,\ \ $\S^{-} =\S_{1}^{d}- \S^{+}$.
First, we show that $R^{-2}I_d^+(R)$ is uniformly bounded in $R$. Indeed,  we
have
$$
|I_d^+(R)|\leq {\frac{|\S_1^d|}{2}} \int_0^1
    \left|(1-u^2)^{-d/2} \exp\left( -\beta R^2\frac{u^2}{1-u^2}\right)-
    1\right|\frac{du}{u},
$$
 and the change of variable
$$s^2 =  \beta R^2\frac{u^2}{1-u^2},$$ leads to
$$
{|I_d^+(R)|\leq  \frac{|\S_1^d|}{2}\int_0^{+\infty}
    \left|\left(1+\frac{s^2}{\beta R^2}\right)^{d/2} \exp (-s^2)-
    1\right|\frac{ds}{s(\frac{s^2}{\beta R^2}+1)}.}
$$
{ A simple computation of the derivative of the function
$$\theta(s) =   1- \left(1+\frac{s^2}{\beta R^2}\right)^{d/2} \exp (-s^2),$$
shows that it is an increasing function on $[0,+\infty[$, for  large enough $R$ ($R>\sqrt{d/2\beta}$).
Therefore,  it is a non-negative function on $[0,+\infty[$, and consequently}
$$
{|I_d^+(R)|\leq  \frac{|\S_1^d|}{2}\int_0^{+\infty}
    \left(\frac{1- \exp(-s^2)}{s}\right)\frac{ds}{\frac{s^2}{\beta R^2}+1}.}
$$
This clearly shows that $R^{-2} I_d^+(R)$ is uniformly bounded for large $R$.

We now focus on the behavior of $I_d^-(R)$ when $R$ goes to infinity
and let $\delta =-\sigma \cdot \omega$ and
\be
\label{Sdelta}
S_\delta(u)= \frac{u(\delta - u)}{1-u^2}.
\ee
Then, we have
\be
\label{Id-}
I_d^-(R)=  \int_{\S^-} d\sigma \int_0^1
    \left[(1-u^2)^{-d/2} \exp\left(\beta R^2 S_\delta(u)\right)-
    1\right]\frac{du}{u}.
\ee
We now show that the dominant part in $I_d^-(R)$ is given by the contribution at
${u=u_{\delta}}$, where ${u_{\delta}}$ is such that
$$\max _{u\in [0,1]} S_{\delta}(u) = S_{\delta} (u_{\delta}).$$
We first have
$$u_\delta=\frac{1}{\delta}\left(1-\sqrt{1-\delta^2}\right), \ \ \ \
 S_{\delta} (u_{\delta})= \frac{1}{2}\left(
 1-\sqrt{1-\delta^2}\right),$$
and observe that
$$S_{\delta}(u) - S_{\delta} (u_{\delta})=- \frac{\left(\delta u -1+\sqrt{1-\delta^2}\right)^2}{2(1-u^2)\left(
 1-\sqrt{1-\delta^2}\right)}.$$
Thus, from \fref{Id-}
\be
\label{Id-2}
\begin{array}{ll}
\ds I_d^-(R)= & \ds  \exp\left(\frac{\beta R^2}{2}\right) \int_{\S^-} d\sigma
 \, \exp\left(-\frac{\beta R^2}{2}\sqrt{1-\delta^2}\right)  \\ & \ds
    \times\int_0^1\left[(1-u^2)^{-d/2} \exp\left(-\beta R^2\frac{\delta \left( u -u_\delta\right)^2}{2u_\delta(1-u^2)} \right)-
   \exp\left(-\frac{\beta R^2}{2}\delta u_\delta
 \right)\right]\frac{du}{u}.
\end{array}
\ee
We now introduce a spherical  system of coordinates (in $\R^d$) for
the integration on the sphere and use identity
$${\int_{\S^-} F(-\sigma\cdot \omega) d\sigma = C_d
\int_{\frac{\pi}{2}}^\pi F(-\cos\theta) (\sin\theta)^{d-2} d\theta  = C_d
\int_{0}^1 F(\delta) (1-\delta^2)^{\frac{d-3}{2}}
 d\delta ,}$$
where $F$ is a real function and $C_d$ is given by \fref{Cd}. In the following, we assume $d\ge 2$ but we have checked by a specific calculation that the results remain valid for $d=1$.
Using this identity in \fref{Id-2},  we  get
\be
\label{Id-3}
\begin{array}{ll}
\ds I_d^-(R)= & \ds  C_d \exp\left(\frac{\beta R^2}{2}\right) \int_{0}^1 d\delta\
 (1-\delta^2)^{\frac{d-3}{2}}\exp\left(-\frac{\beta R^2}{2}\sqrt{1-\delta^2}\right)  \\ & \ds
    {\times\int_0^1}\left[(1-u^2)^{-d/2} \exp\left(-\beta R^2\frac{\delta \left( u -u_\delta\right)^2}{2u_\delta(1-u^2)} \right)-
   \exp\left(-\frac{\beta R^2}{2}\delta u_\delta
 \right)\right]\frac{du}{u}.
\end{array}
\ee
We now perform the following change of variables in \fref{Id-3}
for  both   $\delta$ and $u$
$$t= \frac{u-u_{\delta}}{\sqrt{1-u^2}} \left(\frac{\beta R^2}{2}
\frac{\delta}{u_\delta}\right)^{1/2},   \ \ \     r=
\left(\frac{\beta R^2}{2}\right)^{1/2} (1-\delta^2)^{1/4}, $$
which is also equivalent to
\be
\label{uetr}
u = u(t,r)= \frac{(a^2-r^4)^{1/2} + t(t^2+2r^2)^{1/2}}{a+r^2+t^2},   \
\ \ \    \delta= \left(1- \frac{r^4}{a^2}\right)^{1/2},
\ee
with $a=\beta  R^2/2.$ After some calculations, we get from \fref{Id-3}
\be
\label{Id-4}
\begin{array}{ll}
\ds & \ds I_d^-(R)=  2  C_d \left(\frac{2}{\beta
 R^2}\right)^{d-\frac{1}{2}} \exp\left(\frac{\beta R^2}{2}\right)
 \int_{0}^{{\sqrt{a}}} dr \, r^{2d-3} \left(1- \frac{r^4}{a^2}\right)^{-1/2}
\int_{-(a-r^2)^{1/2}}^{+\infty} dt \\ & \ds
    \times\left[(1-u(t,r)^2)^{-d/2} \exp(-r^2- t^2)-
   \exp\left(-a\right)\right]\frac{(1-u(t,r)^2)^{3/2}}{\left(1+\frac{r^2}{a}\right)^{1/2}u(t,r)
 (1-u(t,r)u_\delta(r))}.
\end{array}
\ee
We now want to analyze the asymptotic behavior of \fref{Id-4} when $a=
\beta R^2/2$ goes to infinity. Thanks to the strong decreasing
properties of $\exp(-t^2-r^2)$, one can expand the terms inside the
integral in the limit  $a\rightarrow +\infty$. To do so, we first obtain from easy calculations
$$ 1-u(t,r)=  \frac{1}{a} \left[ r^2+t^2- t(t^2+2r^2)^{1/2}\right] +
O\left(\frac{1}{a^2}\right) =  \frac{1}{2a} \left[(t^2+2r^2)^{1/2}-t\right]^2 +
O\left(\frac{1}{a^2}\right),$$
and
$$ 1-u(t,r)u_{\delta}(r)=   \frac{1}{a} (t^2+2r^2)^{1/2}\left[(t^2+2r^2)^{1/2}-t\right]+
O\left(\frac{1}{a^2}\right).$$
Then, we plug these relations into \fref{Id-4} to obtain
\be
\label{Id-5}
\begin{array}{ll}
\ds I_d^-(R)\ \sim \   2  C_d \left(\frac{\beta
 R^2}{2}\right)^{-d/2} & \ds\exp\left(\frac{\beta R^2}{2}\right)
 \int_{0}^{+\infty} {dr}\int_{-\infty}^{+\infty}{dt}   \\  & \ds
 \times   r^{2d-3}(t^2+2r^2)^{-1/2}
 \exp(-r^2- t^2) \left[(t^2+2r^2)^{1/2}-t\right]^{2-d},
    \end{array}
\ee
when $a=\beta R^2/2$ goes to infinity.
To  simplify expression  \fref{Id-5} we perform the cylindrical change
 of variable: $t=\rho \cos\phi, $ $r=\rho \sin\phi$, with $\phi \in
 [0,\pi]$,  and get
$$\begin{array}{ll}
\ds I_d^-(R)\ \sim \   2^{3-d}  C_d \left(\frac{\beta
 R^2}{2}\right)^{-d/2} \exp\left(\frac{\beta R^2}{2}\right) & \ds
 \int_{0}^{+\infty}\rho^{d-1}\exp(-\rho^2)\, d\rho   \\  & \ds
 \times\int_{0}^{\pi}   \frac{ \left[ (1+\sin^2\phi)^{1/2} + \cos
 \phi\right]^{d-2}}{(1+\sin^2\phi)^{1/2}} {\sin}\phi \, d\phi.
\end{array}$$
Performing the change of variable $u=\cos\phi$, we obtain
$$\begin{array}{ll}
\ds I_d^-(R)\ \sim \   2^{2-d}  C_d \left(\frac{\beta
 R^2}{2}\right)^{-d/2} \exp\left(\frac{\beta R^2}{2}\right) & \ds
 \Gamma\left (\frac{d}{2}\right )
 \int_{-1}^{+1}   \frac{ \left[ (2-u^2)^{1/2}+u\right]^{d-2}}{(2-u^2)^{1/2}} \, du.
\end{array}$$
Performing the change of variable
 $u=\sqrt{2} \sin t$ in the integral over $u$ we get
 $$\int_{-1}^{1}\frac{\left[ (2-u^2)^{1/2} + u\right]^{d-2}}{(2-u^2)^{1/2}}
 du=2^{d-2}\int_{-\frac{\pi}{4}}^{\frac{\pi}{4}} \sin^{d-2}\left (t+\frac{\pi}{4}\right )\, dt= 2^{d-3} \int_0^\pi (\sin \phi)^{d-2}d\phi,$$
where we have set $\phi=t+\pi/4$ in the last integral. Using  \fref{Cd}, we obtain
$$\begin{array}{ll}
\ds I_d^-(R)\ \sim \   \frac{1}{2} |\S_1^d| \left(\frac{\beta
 R^2}{2}\right)^{-d/2} \exp\left(\frac{\beta R^2}{2}\right) & \ds
 \Gamma\left (\frac{d}{2}\right ).
\end{array}$$
Therefore, $I_d^-(R)$ increases exponentially rapidly while $I_d^+(R)$ increases less rapidly than $R^2$ and can therefore be neglected.
Using \fref{eq5}, we finally obtain \fref{lambda-asymptotique}. This ends the proof of proposition \ref{prop4}.

\section{Asymptotic expressions of the fundamental eigenfunction and eigenvalue}
\label{sec_comp}

In this Appendix, we show that the asymptotic result (\ref{lambda-asymptotique}) can be
directly obtained by a perturbative expansion of the solutions of the
fundamental eigenvalue equation in powers of $\lambda$ in the limit
$\lambda\ll 1$ corresponding to $V\rightarrow +\infty$ (in this Appendix, we introduce more  physical notations and set $v=|v|\in \R^+$ and $V=R\in \R^+$). This method
allows us to treat more general situations, e.g. Fokker-Planck equations with an
arbitrary potential $U(v)$ and an arbitrary diffusion coefficient
$D(v)$. For the Fokker-Planck equation (\ref{FP}) with $U(v)=v^2/2$ and $D=1$, we recover (\ref{lambda-asymptotique}).

For isotropic distributions $f(v,t)$, we consider the Fokker-Planck equation
\begin{eqnarray}
{\partial f\over\partial t}={1\over v^{d-1}}{\partial\over\partial v}\biggl \lbrace v^{d-1} D(v)\biggl ({\partial f\over\partial v}+\beta f {\partial U\over\partial v}\biggr )\biggr\rbrace,\label{mf1}
\end{eqnarray}
where $D(v)$ and $U(v)$ are arbitrary functions of the velocity. We consider the fundamental eigenmode
\begin{eqnarray}
f(v,t)=A e^{\lambda t}g(v),\label{mf2}
\end{eqnarray}
where $A$ is a normalization constant.  We can impose $g(0)=1$ without loss of generality. On the other hand, $g'(0)=0$ and $g(V)=0$. Substituting (\ref{mf2}) in (\ref{mf1}), we obtain the eigenvalue equation
\begin{eqnarray}
\lambda g={1\over v^{d-1}}{d\over d v}\biggl \lbrace v^{d-1} D(v)\biggl ({d g\over d v}+\beta g {d U\over\partial v}\biggr )\biggr\rbrace.\label{mf3}
\end{eqnarray}
It can be integrated into
\begin{eqnarray}
{d g\over d v}+\beta g {d U\over\partial v}={\lambda\over v^{d-1} D(v)}\int_{0}^{v}g(w)w^{d-1}dw.\label{mf4}
\end{eqnarray}
For an unlimited range of velocities ($V\rightarrow +\infty$), the Fokker-Planck equation (\ref{mf1}) admits a steady solution
\begin{eqnarray}
f(v)=\frac{1}{Z}e^{-\beta U(v)},\label{mf5}
\end{eqnarray}
implying that the fundamental eigenvalue is $\lambda(+\infty)=0$. Now for
$V<+\infty$, the eigenvalue $\lambda(V)<0$. However, for $V\gg 1$,
$\lambda(V)\rightarrow 0^-$ and we can
formally expand the solution of the differential equation (\ref{mf3}) in the form
\begin{eqnarray}
g=g_{0}(v)+\lambda g_{1}(v)+\lambda^{2} g_{2}(v)+...\label{mf6}
\end{eqnarray}
To zeroth order, we have
\begin{eqnarray}
{d g_{0}\over d v}+\beta g_{0} {d U\over\partial v}=0.\label{mf7}
\end{eqnarray}
To first order in $\lambda$, we get
\begin{eqnarray}
{d g_{1}\over d v}+\beta g_{1} {d U\over\partial v}={1\over  v^{d-1} D(v)}\int_{0}^{v}g_{0}(w)w^{d-1}dw.\label{mf8}
\end{eqnarray}
With the boundary condition $g_{0}(0)=1$, the first equation can be
integrated in
\begin{eqnarray}
g_{0}(v)=e^{-\beta (U(v)-U(0))}.\label{mf9}
\end{eqnarray}
Substituting this expression in \fref{mf8} we obtain
\begin{eqnarray}
{d g_{1}\over d v}+\beta g_{1} {d U\over\partial v}={e^{\beta U(0)}\over  v^{d-1} D(v)}\int_{0}^{v}e^{-\beta U(w)}w^{d-1}dw,\label{mf9b}
\end{eqnarray}
that we shall solve with the boundary conditions
$g_{1}(0)=g'_{1}(0)=0$. We obtain
\begin{eqnarray}
g_{1}(v)=\chi(v)e^{-\beta (U(v)-U(0))},\label{mf10}
\end{eqnarray}
where $\chi(v)$ is the function defined by
\begin{eqnarray}
\chi'(v)={e^{\beta U(v)}\over v^{d-1} D(v)}\int_{0}^{v}e^{-\beta U(w)}w^{d-1}dw,  \label{mf11}
\end{eqnarray}
with $\chi(0)=0$. To first order in $\lambda$, the fundamental eigenfunction of the Fokker-Planck equation  \fref{mf1} can be written
\begin{eqnarray}
g(v)=e^{-\beta (U(v)-U(0))}\left\lbrack 1+\lambda\chi(v)\right\rbrack.  \label{mf12}
\end{eqnarray}
Using the boundary condition $g(V)=0$, we find that the fundamental eigenvalue is given by
\begin{eqnarray}
\lambda(V)\sim -{1\over \chi(V)}\qquad (V\rightarrow +\infty).\label{mf13}
\end{eqnarray}
This is the main result of this Appendix.

If we specialize on a quadratic potential $U(v)=v^2/2$ (implying a linear friction like in \fref{FP}),
we have
\begin{eqnarray}
g(v)=e^{-\beta v^2/2}\left \lbrack 1+{\lambda}\chi(v)\right \rbrack,  \label{mf14}
\end{eqnarray}
and the function $\chi(v)$ can be written more explicitly
\begin{eqnarray}
\chi'(v)={e^{\beta {v^{2}\over 2}}\over v^{d-1}D(v)}\int_{0}^{v}e^{-\beta {w^{2}\over 2}}w^{d-1}dw,  \label{mf15}
\end{eqnarray}
with $\chi(0)=0$. For large $v$, we have
\begin{eqnarray}
\chi(v)&\sim& \frac{1}{2}\Gamma\left (\frac{d}{2}\right ) \left (\frac{2}{\beta}\right )^{d/2}\int_{0}^{v} e^{\beta {w^{2}\over 2}}{1\over w^{d-1}D(w)}dw , \nonumber\\
&\sim& \frac{1}{4}\Gamma\left (\frac{d}{2}\right ) \left (\frac{2}{\beta}\right )^{d/2+1} {e^{\beta {v^{2}\over 2}}\over  D(v)v^d}. \label{mf16}
\end{eqnarray}
Therefore, for $V\rightarrow +\infty$, the eigenvalue $\lambda(V)$ is given by
\begin{eqnarray}
\lambda(V)\sim -\frac{2\beta}{\Gamma\left (\frac{d}{2}\right )} \left (\frac{\beta V^2}{2}\right )^{d/2} D(V) e^{-\beta {V^{2}\over 2}},  \label{mf17}
\end{eqnarray}
which coincides
 \footnote{On a strict mathematical point of view, the approach presented in this Appendix  is completely formal since we have used a perturbative  approach based on a formal asymptotic expansion. By contrast, the approach developed in Appendix \ref{sec_a3}  leads to an exact formula of $\lambda(V)$ in the case $D=1$  obtained by a rigorous mathematical proof, and the asymptotic behavior (\ref{lambda-asymptotique}) is a  consequence of this exact formula.} with (\ref{lambda-asymptotique})  for $D=1$.

Let us give particular examples:

(i) {\it The King model \cite{king}}: basically, the collisional evolution of stellar systems is described by the gravitational Landau equation (\ref{Landau}) in $d=3$ dimensions. Making a thermal bath approximation, i.e. replacing $f(v_*)$ by the Maxwellian distribution, and assuming that the distribution is spherically symmetric we obtain the Fokker-Planck equation (\ref{FP}) with a diffusion coefficient (see, e.g., \cite{epjb}):
\begin{eqnarray}
D(v)=\frac{K}{v^3}\int_0^v w^2 e^{-\beta \frac{w^2}{2}}\, dv, \label{dfsisd}
\end{eqnarray}
where $K$ is a constant determining the relaxation time in the system. For the  Fokker-Planck equation (\ref{FP}) in $d=3$ with the diffusion coefficient (\ref{dfsisd}), the function $\chi(v)$ is simply given by
\begin{eqnarray}
\chi'(v)={v\over K} \ e^{\beta {v^{2}\over 2}}, \label{mf18}
\end{eqnarray}
with $\chi(0)=0$ (we can appreciate the fortuitous cancelation of the integral first realized by King \cite{king}). This equation is explicitly integrated and we obtain
\begin{eqnarray}
\chi(v)={1\over K\beta} (e^{\beta {v^{2}\over 2}}-1).  \label{mf19}
\end{eqnarray}
The asymptotic evaporation rate \fref{mf13} is then given by
\begin{eqnarray}
\lambda(V)=-{K\beta\over e^{\beta {V^{2}\over 2}}-1}\sim -K\beta e^{-\beta {V^{2}\over 2}}.  \label{mf20}
\end{eqnarray}
The fundamental eigenfunction \fref{mf14}  is the King  solution
\begin{eqnarray}
g(v)=\frac{e^{-\beta {v^{2}\over 2}}-e^{-\beta {V^{2}\over 2}}}{1-e^{- \beta {V^{2}\over 2}}}. \label{mf21}
\end{eqnarray}
This is a lowered isothermal (Maxwell) distribution which vanishes at the escape velocity.
For the fundamental mode, we then have
\begin{eqnarray}
f(v,t)=Ae^{-|\lambda| t}\left (e^{-\beta {v^{2}\over 2}}-e^{-\beta {V^{2}\over 2}}\right ),  \label{mf22}
\end{eqnarray}
where $A$ is a normalization constant. Since the evaporation rate $\lambda$ is small, the system is in a quasi-stationary state given by the fundamental eigenfunction (King model). The total distribution slowly changes in amplitude (without change in form) as stars leave the system. Note that an extension of the King model to the case of fermions (or for the Lynden-Bell \cite{lb} theory of violent relaxation) has been obtained by Chavanis \cite{kingfermi}.

(i) {\it The vortex model \cite{vortex}}: in a cosmogonical context, it has been proposed \cite{bs,bracco} that large-scale vortices may have spontaneously emerged in the protoplanetary nebula, and have captured and accumulated dust particles, leading ultimately to planetesimals and  planets. Chavanis \cite{vortex} has used a Fokker-Planck approach to estimate the evaporation of dust from the vortices due to turbulence (with the final result that evaporation is negligible for relevant particles' sizes). This model is based on a Fokker-Planck equation of the form \fref{FP} (where $v$ plays the role of the position $x$)  with a constant diffusion coefficient $D$ in $d=2$ dimensions. In that case, the function $\chi(v)$ is given by
\begin{eqnarray}
\chi'(v)={e^{\beta \frac{v^{2}}{2}}\over D v}\int_{0}^{v}e^{-\beta \frac{w^{2}}{2}}w \, dw,  \label{mf23}
\end{eqnarray}
with $\chi(0)=0$. This is easily integrated in
\begin{eqnarray}
\chi'(v)=\frac{1}{\beta D v}\left (e^{\beta \frac{v^{2}}{2}}-1\right ),  \label{mf24}
\end{eqnarray}
leading to
\begin{eqnarray}
\chi(v)=\frac{1}{\beta D}\int_{0}^{v}\frac{1}{w}\left (e^{\beta \frac{w^{2}}{2}}-1\right )\, dw.  \label{mf24b}
\end{eqnarray}
This can be rewritten in the form of a series as
\begin{eqnarray}
\chi(v)=\frac{1}{2\beta D}\sum_{n=1}^{+\infty}\frac{(\beta v^{2}/2)^{n}}{n!n}.  \label{mf25}
\end{eqnarray}
This can also be written in the form
\begin{eqnarray}
\chi(v)=\frac{1}{2\beta D}\left\lbrace E_{i}\left (\beta\frac{v^2}{2}\right )-\gamma-\ln\left (\beta\frac{v^2}{2}\right )\right\rbrace,  \label{mf26}
\end{eqnarray}
where
\begin{eqnarray}
E_{i}(x)=\int_{-\infty}^{x}\frac{e^t}{t}\, dt
\label{mf27}
\end{eqnarray}
is the exponential integral and $\gamma$ is the Euler
constant. We see that
\begin{eqnarray}
\chi(v)\sim \frac{1}{\beta D}\int_{1}^{v}\frac{1}{w}e^{\beta \frac{w^{2}}{2}}\, dw,
\end{eqnarray}
for $v$ large.
The asymptotic evaporation rate \fref{mf13}  is then
%\begin{eqnarray}
%\lambda(V)\sim \frac{-2\beta D}{E_{i}\left (\beta\frac{V^2}{2}\right )-\gamma-\ln\left (\beta\frac{V^2}{2}\right )}\sim -D\beta^2 V^2 e^{-\beta V^2/2}.
%  \label{mf28}
%\end{eqnarray}

\begin{eqnarray}
\lambda(V)\sim -D\beta^2 V^2 e^{-\beta V^2/2}.
  \label{mf28}
\end{eqnarray}

\section{Simpler expressions of the function $G(v,u)$}
\label{sec_Gsimple}

Introducing a spherical system of coordinates, the function $G(v,u)$ defined by \fref{G} can be written
\begin{eqnarray}
G(v,u)=\frac{1}{(2\pi)^d}\left (\frac{2\pi\beta}{1-u^2}\right )^{d/2}e^{-\frac{\beta(R^2u^2+v^2)}{2(1-u^2)}}\frac{1}{\int_{0}^{\pi}(\sin\theta)^{d-2}\, d\theta}\int_{0}^{\pi}e^{-\frac{\beta Ru|v|\cos\theta}{1-u^2}}(\sin\theta)^{d-2}\, d\theta.\nonumber\\
\label{fg1}
\end{eqnarray}
Using the identities
\begin{eqnarray}
\int_{0}^{\pi}(\sin\theta)^{d-2}\, d\theta=\frac{\sqrt{\pi}\Gamma\left (\frac{d-1}{2}\right )}{\Gamma\left (\frac{d}{2}\right )},
\label{fg2}
\end{eqnarray}
\begin{eqnarray}
\int_{0}^{\pi}e^{-x\cos\theta}(\sin\theta)^{d-2}\, d\theta=\left (\frac{2}{x}\right)^{\frac{d-2}{2}}\Gamma\left (\frac{d-1}{2}\right )\sqrt{\pi}I_{\frac{d}{2}-1}(x),
\label{fg3}
\end{eqnarray}
we obtain
\begin{eqnarray}
G(v,u)=\Gamma\left (\frac{d}{2}\right )\frac{1}{(2\pi)^{d/2}}\left (\frac{2}{R u |v|}\right )^{\frac{d-2}{2}}\frac{\beta}{1-u^2}e^{-\frac{\beta(R^2u^2+v^2)}{2(1-u^2)}}I_{\frac{d}{2}-1}\left (\frac{\beta R u |v|}{1-u^2}\right ).
\label{fg4}
\end{eqnarray}
In particular, we get
\begin{eqnarray}
G(R\omega,u)=\Gamma\left (\frac{d}{2}\right )\frac{1}{(2\pi)^{d/2}}\left (\frac{2}{R^2 u}\right )^{\frac{d-2}{2}}\frac{\beta}{1-u^2}e^{-\frac{\beta R^2(1+u^2)}{2(1-u^2)}}I_{\frac{d}{2}-1}\left (\frac{\beta R^2 u}{1-u^2}\right ).
\label{fg5}
\end{eqnarray}
We also recall that
\begin{eqnarray}
G(v,0)=\left (\frac{\beta}{2\pi}\right )^{d/2}e^{-\frac{\beta v^2}{2}}.
\label{fg6}
\end{eqnarray}

Let us now consider particular dimensions of space $d=1,2,3$ where the
expression \fref{fg4} can be further simplified.

$\bullet$ In $d=3$, using the identity 
\begin{eqnarray}
I_{1/2}(x)=\sqrt{\frac{2}{\pi x}}\sinh(x),
\label{fg7}
\end{eqnarray}
or  directly integrating \fref{fg1}, we obtain 
\begin{eqnarray}
G(v,u)=\frac{1}{(2\pi)^{3/2}}\frac{1}{R u |v|}\sqrt{\frac{\beta}{1-u^2}}e^{-\frac{\beta(R^2u^2+v^2)}{2(1-u^2)}}\sinh\left (\frac{\beta R u |v|}{1-u^2}\right ).
\label{fg8}
\end{eqnarray}
This can also be written
\begin{eqnarray}
G(v,u)=\frac{1}{2(2\pi)^{3/2}}\frac{1}{R u |v|}\sqrt{\frac{\beta}{1-u^2}}\left\lbrack e^{-\frac{\beta(R u-|v|)^2}{2(1-u^2)}}-e^{-\frac{\beta(R u+|v|)^2}{2(1-u^2)}}\right\rbrack.
\label{fg9}
\end{eqnarray}
In particular,
\begin{eqnarray}
G(R\omega,u)=\frac{1}{2(2\pi)^{3/2}}\frac{1}{R^2 u}\sqrt{\frac{\beta}{1-u^2}}\left\lbrack e^{-\frac{\beta R^2 (1-u)}{2(1+u)}}-e^{-\frac{\beta R^2 (1+u)}{2(1-u)}}\right\rbrack.
\label{fg10}
\end{eqnarray}

$\bullet$ In $d=2$, we obtain 
\begin{eqnarray}
G(v,u)=\frac{1}{2\pi}\frac{\beta}{1-u^2}e^{-\frac{\beta(R^2u^2+v^2)}{2(1-u^2)}}I_{0}\left (\frac{\beta R u |v|}{1-u^2}\right ).
\label{fg11}
\end{eqnarray}

$\bullet$ In $d=1$, using the identity 
\begin{eqnarray}
I_{-1/2}(x)=\sqrt{\frac{2}{\pi x}}\cosh(x),
\label{fg12}
\end{eqnarray}
or directly integrating \fref{fg1}, we obtain 
\begin{eqnarray}
G(v,u)=\frac{1}{(2\pi)^{1/2}}\sqrt{\frac{\beta}{1-u^2}}e^{-\frac{\beta(R^2u^2+v^2)}{2(1-u^2)}}\cosh\left (\frac{\beta R u |v|}{1-u^2}\right ).
\label{fg13}
\end{eqnarray}
This can also be written
\begin{eqnarray}
G(v,u)=\frac{1}{2(2\pi)^{1/2}}\sqrt{\frac{\beta}{1-u^2}}\left\lbrack e^{-\frac{\beta(R u-v)^2}{2(1-u^2)}}+e^{-\frac{\beta(R u+v)^2}{2(1-u^2)}}\right\rbrack.
\label{fg14}
\end{eqnarray}
In particular,
\begin{eqnarray}
G(R\omega,u)=\frac{1}{2(2\pi)^{1/2}}\sqrt{\frac{\beta}{1-u^2}}\left\lbrack e^{-\frac{\beta R^2 (1-u)}{2(1+u)}}+e^{-\frac{\beta R^2 (1+u)}{2(1-u)}}\right\rbrack.
\label{fg15}
\end{eqnarray}

Finally, we note that for $R\rightarrow +\infty$, the eigenfunction \fref{eigenfunction}  is given by
\begin{eqnarray}
f_{\lambda}(v)\simeq M_{\lambda} \left (\frac{\beta}{2\pi}\right )^{d/2}e^{-\beta\frac{v^2}{2}}\left\lbrace 1+\frac{\lambda}{\beta}\int_{0}^{1}\left\lbrack \frac{G(v,u)}{G(v,0)}-1\right\rbrack \frac{du}{u}\right\rbrace.
\label{fg16}
\end{eqnarray}

\section{Recurrence relations for the  moments}

Let us introduce the moments of the distribution function
$$M_n(t)= \int_{B_R} f(v,t) |v|^{2n} dv, \ \ \ n\in \N$$
and integrate Eq. \fref{FP} with $D=1$ against $|v|^{2n}, n\geq 1$. After easy computations,
one gets
$$M_n'(t)= R^{2n}  \int_{\S_R^d} \nabla f \cdot n \, d\sigma - 2\beta n
M_n(t) + 2n (2n+d-2) M_{n-1}(t).$$
Doing the same for $n=0$, we obtain
$$M_0'(t)= \int_{\S_R^d} \nabla f \cdot n \, d\sigma,$$
so that the foregoing expression can be rewritten
$$M_n'(t)= R^{2n} M_0'(t) - 2\beta n
M_n(t) + 2n (2n+d-2) M_{n-1}(t), \ \ \ \ n\geq 1. $$

\end{document}